\documentclass[preprint2]{aastex}
\usepackage {amsmath}
\usepackage {amssymb}
\usepackage {amsfonts}
\usepackage {amsthm}
\usepackage {mathrsfs}
\usepackage {natbib}
\usepackage {latexsym}
\usepackage {graphicx}
\usepackage {dsfont}
\usepackage {times}
\usepackage {txfonts}
\usepackage {rotating}
\usepackage {wasysym}
\usepackage {multirow}
\usepackage {hhline}
\usepackage {hyperref}
\usepackage {color}
\usepackage {bm}
\usepackage{appendix}
\usepackage{acronym}
\usepackage {url}
\usepackage{subfigure}
\usepackage[normalem]{ulem}   

\hypersetup{
  colorlinks=true,        
  linkcolor=blue,         
  citecolor=cyan,         
}

\newcommand{\EMdat}{\bm{S}}
\newcommand{\GWdat}{\bm{D}}
\newcommand{\com}{\bm{\gamma}}
\newcommand{\GWnon}{\bm{\theta}}
\newcommand{\GWfull}{\bm{\Theta}}
\newcommand{\EMnon}{\bm{\phi}}
\newcommand{\EMfull}{\bm{\Phi}}

\newcommand{\Lgal}{L_{\text{g}}}

\slugcomment{Target Astrophysical Journal}

\shorttitle{GW host galaxy identification}
\shortauthors{Fan et al.}

\begin{document}


\title{A Bayesian approach to multi-messenger astronomy:
  Identification of gravitational-wave host galaxies}


\author{XiLong~Fan$^{1,2,}$\altaffilmark{3}, Christopher~Messenger$^{2}$,
  and Ik Siong~Heng$^{2}$}
\affil{1. School of Physics and Electronics Information, Hubei University of Education, 430205 Wuhan, China,\\
2. SUPA, School of Physics and Astronomy, University of Glasgow, Glasgow, G12 8QQ, United Kingdom}


\altaffiltext{3}{Royal Society Newton Fellow, Xilong.Fan@glasgow.ac.uk}



\begin{abstract}
  We present a general framework for incorporating astrophysical
  information into Bayesian parameter estimation techniques used by
  gravitational wave data analysis to facilitate multi-messenger
  astronomy.  Since the progenitors of transient gravitational wave
  events, such as compact binary coalescences, are likely to be
  associated with a host galaxy, improvements to the source sky
  location estimates through the use of host galaxy information are
  explored.  To demonstrate how host galaxy properties can be
  included, we simulate a population of compact binary coalescences
  and show that for ${\sim}8.5\%$ of simulations with in $200$Mpc, the
  top ten most likely galaxies account for a ${\sim}50\%$ of the total
  probability of hosting a gravitational wave source. The true
  gravitational wave source host galaxy is in the top ten galaxy
  candidates ${\sim}10\%$ of the time.  Furthermore, we show that by
  including host galaxy information, a better estimate of the
  inclination angle of a compact binary gravitational wave source can
  be obtained. We also demonstrate the flexibility of our method by
  incorporating the use of either B or K band into our analysis.
\end{abstract}





\acrodef{GW}[GW]{gravitational-wave}
\acrodef{BNS}[BNS]{binary neutron star}
\acrodef{CBC}[CBC]{Compact binary coalescence}
\acrodef{HMNS}[HMNS]{hypermassive neutron star}
\acrodef{SGRB}[SGRB]{short-duration gamma-ray burst}
\acrodef{LGRB}[LGRB]{long-duration gamma-ray burst}
\acrodef{GRB}[GRB]{gamma-ray burst}
\acrodef{ET}[ET]{Einstein Telescope}
\acrodef{NS}[NS]{neutron star}
\acrodef{EM}[EM]{electromagnetic}
\acrodef{SNR}[SNR]{signal-to-noise ratio}
\acrodef{PDF}[PDF]{probability distribution function}
\acrodef{EOS}[EOS]{equation of state}
\acrodef{GWGC}[GWGC]{Gravitational Wave Galaxy Catalogue}
\acrodef{UNGC}[UNGC]{Updated Nearby Galaxy Catalog}
\acrodef{MMPF}[MMPF]{multi-messenger prior function}

\section{Introduction}\label{sec:introduction}

%
%
The first detection of \acp{GW} will herald the dawn of gravitational
wave astronomy and will provide a new way of exploring our universe
complimenting existing \ac{EM} observations.  With gravity coupling
very weakly to matter, the detection of gravitational waves is an
immense ongoing challenge that pushes both technological and
scientific boundaries.  Advanced LIGO~\citep{Harry:2010} and Advanced
Virgo~\citep{Virgo:2009} are expected to have sensitivities that make
the detection of \acp{GW} a very real prospect in the next few years.
Sources of \acp{GW} can be classed into 4 broad categories. Continuous
\ac{GW} sources, such as rapidly rotating neutron stars, emit
quasi-sinusoidal \acp{GW} over durations much longer than the lifetime
of the detectors. Stochastic \acp{GW} can take the form of a
cosmological background, analogous to the \ac{EM} cosmic microwave
background, or could arise from a cacophony of \ac{GW} sources at
closer distances. Burst \acp{GW} are transient signals with poorly
modelled or unknown waveforms. Examples of burst sources are
supernovae and the merger and post-merger phases of merging compact
binaries. \acp{CBC} are inspiralling binary systems where either or
both constituents are a black hole or neutron star. \acp{CBC} are the
best characterised and one of the most promising sources for the
Advanced detectors, with a realistic expected rate of 20 such events
per year observed at design sensitivity~\citep{Collaboration:2013ua}.

%
\emph{Multi-messenger astronomy} involves the joint observation of
astrophysical phenomena using a combination of \ac{EM}, neutrinos or
\ac{GW} observatories. Examples of multi-messenger astronomy involving
\acp{GW} include gamma-ray burst observations by
Swift~\citep{Evans2012ApJS..203...28E,2012ApJ...759...22K,2013PhRvD..88l2004A}
and Fermi~\citep{2013arXiv1303.2174B} and all satellite-based
gamma-ray experiments \citep{2012ApJ...760...12A,2010ApJ...715.1438A},
optical transients by several telescopes \citep{2014ApJS..211....7A}
(see a general implementation in \cite{2012A&A...539A.124L}) and
astroparticle observatories \citep[e.g. high-energy
neutrinos,][]{2011PhRvL.107y1101B,AndoRevModPhys.85.1401}.  Such joint
observations are likely to be mutually beneficial. For \ac{GW}
observations, an observation in the \ac{EM} spectrum will allow
\ac{GW} data analysts to focus their searches on a reduced parameter
space, thereby improving the sensitivity of their
analyses. Conversely, the detection of a \ac{GW} signal can trigger
\ac{EM} observatories to search for counterpart signals in their
respective observation bands. Joint observations will also allow for
better characterisation of the signal progenitor, and a richer
interpretation of the results of the \ac{GW} search. For example,
searches for \acp{GW} in association with GRB051103
\citep{2012ApJ...755....2A} and GRB070201 \citep{2008ApJ...681.1419A}
have ruled out the possibility that their progenitors are \ac{CBC}
sources in nearby galaxies.  In addition to improving sky location
estimates, identifying the host galaxies of \ac{GW} signal progenitors
will enrich this observation by allowing the progenitor environment to
be studied which would, for example, provide insight into the
evolution of \ac{CBC} systems.  Searches for \ac{GW} signatures from
isolated neutron stars, of both continuous and transient natures, are
informed by radio and X-ray observations \citep[see, for example,
pulsar glitch and continuous \ac{GW}
searches,][]{2007PhRvD..76d3003C,2008CQGra..25r4016H}.  It is also
possible to exploit \acp{GW} from \ac{CBC} sources to obtain
luminosity distance estimates for source progenitors whilst EM
observations of the same event (e.g. gamma-ray bursts) will provide
redshift information which can be used to measure the Hubble constant
\citep[e.g.][]{Schutz1986Natur.323..310S,2012PhRvD..86d3011D}.

%
Source sky localisation is one of the crucial ingredients for
multi-messenger astronomy
{\citep[e.g.][]{2003ApJ...591.1152S,2008CQGra..25k4038S,Wen2008JPhCS.122a2038W,Wen2010PhRvD..81h2001W,Veitch2010PhRvD..81f2003V,2011CQGra..28j5021F}}. Identifying
that a \ac{GW} signal originates from a sky location consistent with
an \ac{EM} counterpart will establish a clear link between the two
observations.  Using \ac{GW} observations to trigger searches for
\ac{EM} counterparts will require accurate and precise estimates of
the source sky location to reduce the areas of the sky which the
searches for the \ac{EM} counterpart are to be performed.
%
%
Uncertainty on the estimated sky location of transient \ac{GW} signals
varies with the strength of the \ac{GW} signal as well as the location
of each \ac{GW} observatory and their orientations relative to one
another.  For example, a 3--detector network consisting of the two
Advanced LIGO observatories and Advanced Virgo will provide sky
localisation estimates of a few tens of square
degrees~\citep{Collaboration:2013ua,2013arXiv1312.6013S,2013arXiv1310.7454G}
which is a significant challenge for many \ac{EM} observatories to
scan in search for an \ac{EM} counterpart.  In the initial years of
Advanced detector operation the sky localisation ability is further
impeded by only having the two LIGO detectors (with a slightly less
sensitive Virgo after $\sim 1$ year of operation).  The corresponding
sky position uncertainties in this case are $\mathcal{O}(100-1000)$s
of square degrees~\citep{2014arXiv1404.5623S}.

%
An observed \ac{GW} signal, in particular from \ac{CBC} sources,
provide constraints on both source distance and sky location.  With
this information, if the host galaxy of the \ac{GW} signal progenitor
can be identified, then \ac{EM} observations can focus only on the
region of sky associated with this galaxy.
Amongst the many existing galaxy catalogues, the
\ac{GWGC}~\citep{White2011CQGra..28h5016W} has been
specifically compiled for current follow-up searches of optical
counterparts from \ac{GW} triggers.  \cite{2010PhRvD..82j2002N}
proposed a ranking statistic to identify the most likely \ac{GW} host
galaxy based on galaxy distance and luminosity and the sky position
error box.  This ranking method has been adopted in optical follow-up
pipeline design~\citep{2013ApJS..209...24N}, and optical follow-up
observation~\citep{2014ApJS..211....7A}. 

While analyses of Burst \ac{GW} signals tend not to provide estimates
on source distances, due to the assumed unmodelled nature of most
burst \ac{GW} progenitors, it is still desirable to identify potential
host galaxies for Burst sources. There are distant-independent
algorithms for associating potential host galaxies with observed Burst
signals, such as assigning a host galaxy probability based on the
surface density of differential number counts of galaxies
\citep[e.g.][]{Bloom2002AJ....123.1111B}.

The expected reach of Advanced LIGO and Advanced Virgo at design
sensitivity is ${\sim}200$ Mpc, which is beyond the current range of
the \ac{GWGC}.  The detection efficiency improvements for wide-field
\ac{EM} follow-ups obtained through the use of galaxy catalogs has
been investigated by~\citep{2014ApJ...784....8H}.  They estimate that
an average of ${\sim}500$ galaxies are located in a typical \ac{GW}
sky location error box for NS/NS mergers with Advanced LIGO
(${\sim}20$ square degrees), up to range of $200$Mpc.  By taking into
account the \ac{GW} measurement error in distance and sky location, it
was found that the use of a complete galaxy catalogue can improve the
probability of successful identification of the host galaxy by
${\sim}10$---$300\%$ (depending on the telescope field-of-view)
relative to follow-up strategies that do not utilize catalogues.

%
In the following section we describe the statistical formalism on
which our Bayesian approach is based.  We define how information from
\ac{EM} observations can be combined with information from \ac{GW}
observations and how this leads to an enhanced ability to identify
\ac{GW} source host galaxies.  We also highlight the additional
inference power that \ac{EM} observations lend to the estimation of
some \ac{GW} source parameters, such as the inclination angle of a
\ac{CBC} event.  We then describe the specific case of galaxy
catalogues representing the \ac{EM} observation in
Sec.~\ref{sec:EM_galaxy}.  In Secs.~\ref{sec:CBC}
and~\ref{sec:implementation} respectively we describe \ac{CBC} signal
waveforms and how, in practice, we combine information from galaxy
catalogues into our analysis of \ac{GW} data and the simulation
details.  In Sec.~\ref{sec:results} we report the results of the
\ac{CBC} simulations that we have performed to validate our method. We
conclude with Sec.~\ref{sec:discussion} with a discussion of our
results.

\section{Inference with joint \ac{EM} and \ac{GW} observations}\label{sec:inference}

Our aim is to define a method for combining the information contained
within EM observations and that obtained through \ac{GW} observations.
In doing so we choose to treat the \ac{EM} and \ac{GW} observations
and analyses separately up until the point at which parameter
estimation on each dataset has been completed.  By this we mean that
the final output of many \ac{EM} observations are represented by
astronomical results which contain both direct measurements (such as
sky position and flux), and estimated values (such as distance and
luminosity).
The final output of the \ac{GW} observation is represented by the
posterior probability density describing the \ac{GW} source
parameters.  Given our model assumption (see Sec.~\ref{sec:EM_galaxy})
that \ac{GW} sources are hosted within galaxies there will be common
parameters between the two observations, namely the sky position and
potentially the distance (dependent upon whether the \ac{GW} source
and/or the \ac{EM} observation include an estimable distance
parameter).

In the following sections we will describe how we combine the
information from both observations to enhance our parameter estimation
on these common parameters.  We also show that due to the correlations
between sets of \ac{GW} parameters, improved knowledge of a parameter
that is common between \ac{EM} and \ac{GW} observations can also
enhance our parameter estimation abilities on non-common parameters.

\subsection{Definitions of relevant quantities}

To facilitate the formulation of the proposed method for joint
\ac{EM}--\ac{GW} observations, we start by formally defining the
relevant quantities within our problem:

\begin{enumerate}
\item The parameter set common to both sets of observations is denoted
  by $\com$.  In practice in most cases this will consist of the
  astrophysical source location parameters, the sky position
  $\alpha,\beta$ and the distance $d$.
\item The complete parameter set governing an individual \ac{GW} event
  is denoted by $\GWfull=(\GWnon,\com)$ which includes the common
  parameter set $\com$ but also the set $\GWnon$ which does not
  influence the \ac{EM} observations.  As an example, for a compact
  binary coalescence source $\GWnon$ could contain (amongst other
  parameters) the chirp mass, $\mathcal{M}$.  The specific choice of
  what constitutes a non-common parameter is dependent upon the
  \ac{GW} source type and the \ac{EM} observation.
\item The complete parameter set governing the entire \ac{EM} dataset
  is denoted by $\EMfull=(\EMnon,\com)$ which includes the common
  parameter set $\com$ but also the set $\EMnon$ which does not
  influence the \ac{GW} observations. As an example, if our \ac{EM}
  dataset is represented by a galaxy catalogue this could include the
  galaxy luminosity $\Lgal$, metallicity $Z$, morphology, etc.
\item The \ac{GW} dataset is denoted by $\GWdat$ potentially
  consisting of the outputs from multiple \ac{GW} detectors such that,
  for the $i^{\text{th}}$ detector, $D_i(t) = h_{i}(t,\GWfull) +
  n_{i}(t)$, where $h_{i}(t,\GWfull)$ is the \ac{GW} signal and
  $n_{i}(t)$ is the noise from the \ac{GW} detector.  The data can
  be defined equivalently in the frequency domain.
\item The \ac{EM} data used is denoted by $\EMdat$. We do not formally
  define the specific constituents of $\EMdat$ other than stating that
  they consist of multiple \ac{EM} observations.
\item We use $M$ to define our underlying model assumption that links
  the \ac{GW} to \ac{EM} observations. When the \ac{EM} observation is
  represented by a galaxy catalogue, then our model assumes \ac{GW}
  sources are hosted within galaxies (see Sec. \ref{sec:EM_galaxy}).
\item Within our Bayesian framework we use the standard $I$ to contain
  all additional information.
\end{enumerate}

\subsection{Combining \ac{GW} and \ac{EM} observations}\label{sec:combining}

Our intention is to compute the posterior distribution of the common
parameter set $\com$ conditional on both datasets $\EMdat$, $\GWdat$
and our underlying model $M$.  Let us start by using Bayes theorem to
express the joint distribution on the complete \ac{GW} parameter set
as
\begin{equation}
\label{eq:joint_pos1}
p(\com,\GWnon |\GWdat,\EMdat,M,I)=\frac{p( \com,\GWnon |\EMdat,M,I) p(\GWdat | \EMdat,\com,\GWnon,M,I)}{p(\GWdat | \EMdat ,M,I)}.
\end{equation}
Taking the first term in the numerator we find that
\begin{align}
  p(\com,\GWnon |\EMdat,M,I) &=
  p(\com|\EMdat,\GWnon,M,I)p(\GWnon|\EMdat,M,I)\nonumber\\
  &=p(\com|\EMdat,M,I)p(\GWnon|I)
\end{align}
where we assume that the common parameters are independent of the
non-common \ac{GW} parameters and that our knowledge of the non-common
\ac{GW} parameters are not informed by the \ac{EM} observations alone.

Here $p( \com|\EMdat,M,I)$ takes the form of a prior but can also be
viewed as the posterior on the common parameters $\com$ as defined by,
for example, a catalogue of host galaxies.  The quantity $p(\GWdat |
\EMdat,\com,\GWnon,M,I)$ is the likelihood of obtaining the dataset
$\GWdat$ given $\EMdat$, $\com$ and $\GWnon$ and $p(\GWdat | \EMdat
,M,I)$ is a normalising factor often referred to as the Bayesian
evidence.

The likelihood term in Eq.~\ref{eq:joint_pos1} can be simplified by
noting that, given $\com$, the probability of measuring $\GWdat$ is
fully specified making $\EMdat$ redundant.  Hence
\begin{equation}\label{eq:first_app}
  p(\GWdat | \EMdat,\com,\GWnon,M,I)\equiv p(\GWdat | \com,\GWnon,M,I).
\end{equation}
We can re-express Eq.~\ref{eq:first_app} via Bayes theorem
to give us
\begin{equation}
  \label{eq:rearrange}
  p(\GWdat|\com,\GWnon,M,I)=\frac{p(\com,\GWnon|\GWdat,M,I) p(\GWdat|I)}{p(\com,\GWnon|I)}
\end{equation}
which we can now substitute back into Eq.~\ref{eq:joint_pos1} to give
\begin{equation}
\label{eq:joint_pos3}
p(\com,\GWnon|\GWdat,\EMdat,M,I) =\frac{p(\GWdat|I) p(\com,\GWnon|\GWdat,M,I) p(\com|\EMdat,M,I)}{p(\GWdat|\EMdat,M,I)p(\com|I)}.
\end{equation}
Taking groups of elements in turn we see that there is a constant
normalising prefactor equal to a ratio of Bayesian evidences. This is
technically a Bayes Factor between two models, the first model being
that the \ac{GW} data contains a \ac{GW} signal and the second model
stating that there is a \ac{GW} signal in the \ac{GW} data and it is
consistent with the \ac{EM} observation.  It is clearly a function of
both \ac{GW} and \ac{EM} datasets but is independent of the
parameters.

The term $p(\com,\GWnon|\GWdat,M,I)$ is the joint posterior
probability distribution on all \ac{GW} parameters obtained from a
\ac{GW}-only analysis.  We have, as a denominator, the prior on the
common parameters uninfluenced by the \ac{EM} or \ac{GW} observations.
The reason for its appearance as a denominator is that we must account
for the fact that this prior has already been used implicitly twice
before, once in constructing the \ac{GW} posterior and we assume also
in the final term, the \ac{EM} posterior.

To obtain our final goal of a posterior distribution on the common
parameters alone we now simply marginalise over the non-common \ac{GW}
parameters in Eq.~\ref{eq:joint_pos1}.  These parameters are
referenced only in the \ac{GW}-only joint posterior term and hence the
posterior distribution on the common parameter set $\com$ conditional
on both datasets $\EMdat$ and $\GWdat$ is
\begin{equation}
\label{eq:joint_pos4}
p(\com|\GWdat,\EMdat,M,I) =\frac{p(\GWdat|I)p(\com|\GWdat,M,I) p(\com|\EMdat,M,I)}{p(\GWdat|\EMdat,M,I)p(\com|I)}.
\end{equation}
Note that again, the common-parameter prior distribution remains as a
denominator.  We address the practical implementation of this feature
in Appendix~\ref{sec:GWsamples}.

\subsection{Enhanced inference on \ac{GW}
  parameters}\label{sec:inclination}

Additional information from \ac{EM} observations can be used to
improve the inference on non-common \ac{GW} signal parameters.  This
ability applies to non-common parameters that exhibit correlation in
the \ac{GW} posterior with one of more parameters in $\com$.  An
example of such a parameter is the inclination angle $\iota$ in
\ac{CBC} sources which is strongly correlated with the distance
parameter $d$.  We stress that this correlation does not exist within
the astrophysically motivated prior distribution and is generated by
the inclusion of information from the \ac{GW} dataset.

Using Eq.~\ref{eq:joint_pos3} it can be seen that in general,
marginalising over the common parameter set as follows
\begin{align}
\label{eq:joint_pos5}
p(\GWnon|\GWdat,\EMdat,M,I) =&\frac{p(\GWdat|I) } {
  p(\GWdat|\EMdat,M,I)}\nonumber\\
&\times\int d\com\frac{p(\com,\GWnon|\GWdat,M,I)
  p(\com|\EMdat,M,I)}{p(\com|I)}
\end{align}
does not return a quantity proportional to $p(\GWnon|\GWdat,M,I)$.  In
this case the \ac{EM} posterior on the common parameters and the prior
in the denominator act to influence the \ac{GW} posterior to specific
regions of the $\com$ space.  If the joint \ac{GW} posterior $\com$ is
correlated to any subset of $\GWnon$ then the inclusion of \ac{EM}
data will have enhanced the inference on these non-common parameters.

In the \ac{CBC} example, if the common parameters include distance
then the \ac{EM} dataset may identify a possible range or ranges of
$d$ that are localised within the \ac{GW}-only inferred ranges.
Having more tightly constrained distance values for a given signal
will then correspondingly improve constraints on the inclination angle.

\section{Galaxy catalogues as \ac{EM} observations}\label{sec:EM_galaxy}

Throughout, we work under the assumption that sources
of \acp{GW} reside within (or in close proximity to) the normal matter
seen as galaxies.  Sky position and distance estimation from \ac{GW}
observations alone are expected to be relatively uncertain.  Hence
galaxy catalogues, which we refer to as the \ac{EM} observation, will
help significantly in identifying the \ac{GW} source host galaxy.

The dependence of the \ac{EM} data on the common parameter set $\com$
appears clear under the assumption of our general model.  If we assume
that our galaxy catalogue is complete then there can be no probability
of a \ac{GW} source at any location \emph{not} coincident with a
galaxy.  We generalise this statement below accounting for the
incompleteness of our galaxy catalogues.  Additionally, our underlying
model may include other parameter dependencies whereby, based on the
catalogue alone, one galaxy would be favoured over another.  The
obvious example parameter in this case is the galaxy luminosity which
is strongly related to the galaxy mass and hence the probability of
hosting a \ac{GW} progenitor. Beyond this example we may consider
further properties of galaxies that would influence our belief in the
presence of a \ac{GW} source at one location as opposed to
another. These may include galaxy type, or metallicity.  For future
\ac{GW} searches sensitive to cosmological distances the redshift will
influence this belief based on stellar evolution.

All of this information must be encoded into what we call the
\ac{MMPF}, $p(\com|\EMdat,M,I)$, where we make clear the inclusion of
$M$ representing our underlying model.  We interpret the galaxy
catalogue location information (sky position and distance) as relating
to the probability of the presence of a galaxy, but {\bf not} that it
is necessarily related to a \ac{GW} event.  To make this clearer, via
Bayes theorem we now decompose this function to give
\begin{equation}
  p(\com|\EMdat,M,I)\propto p(\com|\EMdat,I)p(M|\com,\EMdat,I).
\end{equation}
allowing us to assign this probability based on our understanding of
\ac{GW} progenitor abundance as a function of \ac{EM} information. The
term $p(M|\com,\EMdat,I)$ represents the probability of our model $M$
(that \ac{GW} sources are hosted by galaxies) given a specific
location and the \ac{EM} data.
The term $p(\com|\EMdat,I)$ relates to the possible spatial location
of the host galaxy and can be inferred directly from the galaxy
catalogue.  This however requires us to address the issue of the
completeness of \ac{EM} dataset $S$.  Note that the \ac{EM} dataset
that we have (in the form of the galaxy catalogue) if used un-modified
will artificially limit the spatial extent of our \ac{MMPF}. Since the
\ac{GWGC} extends only to $100$ Mpc compared with the $200$ Mpc
sensitive range of the Advanced \ac{GW} detectors this is certainly
the case. 

%
We therefore use an approximation to $p(\com|\EMdat,I)$ that encodes
our ignorance of the galaxy distribution beyond $100$ Mpc. In this
region $d>100$Mpc we assume a uniform galaxy distribution in volume
and hence model the distance prior as $\propto d^{2}$.  We therefore
also use an isotropic prior on sky position giving us constant priors
on the right ascension and the cosine of the declination. This
approximate approach assumes an abrupt transition from the galaxy
catalogue prior to spatial ignorance at $100$ Mpc and beyond.  We
therefore implicitly assume $100\%$ completeness of the galaxy
catalogue up to $100$Mpc. This assumption is incorrect since we know
that, the \ac{GWGC} for example, is only $60\%$ complete beyond $60$
Mpc.  However, this is a small effect compared to the bias we avoid by
taking into account our ignorance beyond $100$ Mpc.

To construct $p(\com|\EMdat,M,I)$, we first consider
$p(\com|\EMdat,I)$ and in doing so we make some assumptions about the
sky location of our desired \ac{GW} signal progenitor with respect to
the galaxy properties contained within $\EMdat$.  To provide a proof-of-principle of our proposed
method, we assume a straightforward form of $p(\com|\EMdat,I)$ such
that
\begin{align}\label{eq:mmpf1}
  p(\com|\EMdat,I) =& \left(\frac{D_{\text{gc}}}{D_{\text{GW}}}\right)^{3}\frac{1}{N}\sum_{j=1}^{N}\delta(\alpha-\alpha_j,\beta-\beta_j,d-d_j)\nonumber \\
  &+ \frac{3}{4\pi D_{\text{GW}}^{3}}H(d-D_{\text{gc}})d^{2}
\end{align}
where $N$ is the number of galaxies in the catalogue and the
subscripted sky position parameters $\alpha_j,\beta_j$ and $d_j$
denote the galaxy catalogue right ascension, declination and
luminosity distance values of the $j$'th galaxy respectively.  We use
$D_{\text{gc}}$ and $D_{\text{GW}}$ to represent the range of the
galaxy catalogue and the sensitive range of the \ac{GW} detector
network respectively and $H$ is a Heaviside step function.
Eq.~\ref{eq:mmpf1} is constructed to satisfy the constraints that
inside the catalogue range only galaxies are considered valid source
locations wheras outside, all locations are valid.  Also, under the
assumption of uniform galaxy distribution in volume the total
probability in each region (within the galaxy catalogue range and
beyond) is proportional to the volume of that region.

The additional function required to complete the \ac{MMPF} is the
probability of a \ac{GW} source existing given a location.  This is
chosen as
\begin{align}\label{eq:mmpf2}
  p(M|\com,\EMdat,I) \propto&
  \sum_{j=1}^{N}\delta(\alpha-\alpha_j,\beta-\beta_j,d-d_j)\mathcal{L}_{Bj}\nonumber\\
  &+ H(d-D_{\text{gc}})\bar{\mathcal{L}}_{B}
\end{align}
where $\mathcal{L}_{Bj}$ is the observed B-band luminosity of the
$j$'th galaxy reported in the catalogue and $\bar{\mathcal{L}}_{B}$ is
the mean B-band luminosity.  As before, the first part of the function
describes the known properties of the galaxies within the catalogue
and the second part represents our ignorance of the galaxies beyond
the catalogue range.  In this latter case we assign a \ac{GW} host
probability proportional to the mean/expected B-band luminosity which
we obtain from the distribution of luminosities from within the
catalogue\footnote{For the \ac{GWGC} catalogue, since we know that it is
${\sim}100\%$ complete to within $60$ Mpc we take the average
luminosity from galaxies within that range.}. The final \ac{MMPF} can
then be written as
\begin{align}\label{eq:mmpf}
  p(\com|\EMdat,M,&I) \propto \nonumber \\
  &\left(\frac{D_{\text{gc}}}{D_{\text{GW}}}\right)^{3}\frac{1}{N}\sum_{j=1}^{N}\delta(\alpha-\alpha_j,\beta-\beta_j,d-d_j)\mathcal{L}_{B_j}\nonumber\\
  &+ \frac{3\bar{\mathcal{L}}_{B}}{4\pi
    D_{\text{GW}}^{3}}H(d-D_{\text{gc}})d^{2}.
\end{align}

This version of \ac{MMPF} is only a first order approximation to a
potentially fully robust way of accounting for catalogue completeness
and other \ac{EM} data effects.  Enhancements of the
proof-of-principle analysis described in this paper can naturally
account for completeness effects due to intrinsic luminosity variation
between galaxies, sky position survey sensitivity, etc.  It can also
allow smooth transitions between our knowledge based on the catalogue
and our ignorance beyond the catalogue range.  In this case the
Heaviside function in Eqns.~\ref{eq:mmpf1}--\ref{eq:mmpf} would be
replaced by a more physically motivated function. One can also account
for the uncertainties in the measurements of $\EMfull$ or potential
offsets in the observed signal from the centre of its host galaxy
(e.g. supernovae kicks) by assigning a distribution to each parameter
with a finite variance covering the uncertainties and offsets instead
of the Dirac Delta functions used in
Eqns.~\ref{eq:mmpf1}--\ref{eq:mmpf}. While such a function is
straightforward to construct mathematically, it will significantly
increase the computation cost of our analysis. We leave this for
future work but note that our simplistic step-function application of
a non-zero \ac{MMPF} beyond the catalogue range accounts the bulk of
the biases imposed by the lack of catalogue completeness.

\section{Implementation method}\label{sec:implementation}
 
In this section we describe how our approach can be applied to a
specific source type in conjunction with our choice of \ac{EM}
observation, a galaxy catalogue. The source we choose is \ac{CBC}
since this allows us to make use the existing lalinference
software~\citep{2013PhRvD..88f2001A} for performing the \ac{GW}
inference component of our analysis.  We discuss the general
application of our method in Sec.~\ref{sec:discussion}.

\subsection{\acp{GW} from compact binary coalesences}\label{sec:CBC}

The \ac{GW} signal from a \ac{CBC} can be divided into 3 parts, the
inspiral, merger and ringdown of the final object.  For ground-based
\ac{GW} detectors the inspiral stage of \acp{BNS} contains the
dominant \ac{SNR} component and the merger and ringdown can be
neglected with regards to detection and sky localisation.  We can
therefore write the inspiral frequency domain waveform using the
stationary phase approximation~\citep{1994PhRvD..49.1707D} as
\begin{equation}
  \tilde{h}(f) =
  \frac{\mathcal{A}(\varphi,\mathcal{M})}{d}f^{-7/6}e^{-i\left\{\Psi(f;\mathcal{M},\eta)-2\pi
  f\left(t_{c}-\frac{\vec{n}(\alpha,\beta)\cdot\vec{r}}{c}\right) - \phi_{c}\right\}}.
\end{equation}
We define the total mass $M = m_{1} + m_{2}$ and the symmetric mass
ratio $\eta = m_{1}m_{2}/M^{2}$ where $m_{1}$ and $m_{2}$ are the
component masses. The chirp mass $\mathcal{M}$ is defined as
$\mathcal{M} = M\eta^{3/5}$ and $d$ is the luminosity distance of the
\ac{GW} source. The quantity $\mathcal{A}(\varphi)$ is a factor that
is determined by the amplitude response of the \ac{GW} detector and is
a function of the chirpmass and the nuisance parameters $\varphi =
(\alpha, \beta, \iota, \psi)$ where $\alpha$ and $\beta$ are the sky
position coordinates and $\iota$ and $\psi$ are the orbital
inclination and \ac{GW} polarization angles respectively.  The phase
$\Psi(f)$ is a function of the frequency and the mass parameters and
in the post-Newtonian point-particle approximation can be Taylor
expanded in powers of the dimensionless quantity $(\pi M
f)^{2/3}$~\citep{2005PhRvD..71h4008A,2005PhRvD..72f9903A}.  There is
also a frequency dependent phase component with argument proportional
to the time of coalescence $t_{c}$ minus a time delay term
representing the \ac{GW} travel time from a common reference point to
the detector location.  This time delay is equal to the dot product
between the detector position vector $\vec{r}$ and the unit sky
position vector $\vec{n}$.  Finally there is a constant phase
component $\phi_{c}$, the phase at coalescence.


The common parameters for this
galaxy-\ac{CBC} joint observation are the sky location and distance,
giving $\com=(\alpha,\beta,d)$.  The remaining \ac{GW} parameters in
our analysis
$\GWnon=(m_1,m_2,\iota,\psi,t_{\text{c}},\phi_{\text{c}})$ are
marginalised over as described in Eq.~\ref{eq:joint_pos4} and in
Appendix~\ref{sec:GWsamples}. Reiterating our underlying model, we
assume that the \ac{GW} source resides within a galaxy and hence,
following the arguments outlined in Sec.~\ref{sec:EM_galaxy} we define
the \ac{MMPF} $p(\com|\EMdat,M,I)$ using Eq.~\ref{eq:mmpf}.

\subsection{Simulation details}\label{sec:implementation}

To demonstrate the effectiveness of the proposed method, we simulate a
population of \ac{GW} signals from \ac{BNS} coalesences.  For
simplicity, each \ac{BNS} coalesence is simulated with the exact sky
position and distance parameters of a randomly chosen galaxy (we
ignore potential supernovae kicks and galaxy catalogue distance
uncertainties).  The galaxy is chosen from a galaxy catalogue
according to an \ac{MMPF} of the form defined by Eq.~\ref{eq:mmpf}. We
perform 3 separate simulations using a combination of 2 different
galaxy catalogues, the initial and advanced LIGO--Virgo \ac{GW}
detector network and different specific choices of luminosity band
used in the \ac{MMPF}.
Table~\ref{simulation_p} gives a summary of our simulations.
Note that the \ac{MMPF} is weighted by B band luminosity for simulations 
in S1 while K band luminosities are used for S2 and S3 simulations.

The \ac{MMPF} defined in Eq.~\ref{eq:mmpf} allows for the probability
of a source being hosted beyond the range of the galaxy catalogue.
Whilst we use this \ac{MMPF} in our analysis of the simulated data,
due to computational constraints, we do not simulate \emph{all}
required signals beyond the galaxy catalogue range.  We perform an
equal number of simulations inside and outside the range where we
should in fact simulate the fraction $\approx 1 -
(D_{\text{gc}}/D_{\text{GW}})^3$ beyond $D_{\text{gc}}$.  To account
for this bias we recycle our costly simulations.  This is done by
taking the \ac{GW} posterior samples for each simulation beyond
$100$Mpc and randomly translating the ensemble of samples in right
ascension to a new sky location.  For a uniform distribution of
sources in volume our \ac{MMPF} demands that there be on average 7
times as many sources between $100$ and $200$Mpc as those within
$100$Mpc.  Therefore, each simulation beyond $100$Mpc is recycled 6
times. This way we are able to efficiently model the statistical
behaviour and variation between these posterior distributions for
moderate computational cost.

The first simulations (S1) assumed a 3 network consisting of the two
Advanced LIGO detectors and Advanced Virgo.  We selected 1000 \ac{GW}
host galaxies from the \ac{GWGC}~\citep{White2011CQGra..28h5016W}
according to the first term in the \ac{MMPF} function
(Eq.~\ref{eq:mmpf}) assuming a complete catalogue up to $100$Mpc.  And
additional 1000 were selected uniformly in volume between
$100$--$200$Mpc according to the second term.  The latter were then
recycled to represent 7000 injections hence closely representing
samples drawn directly from the \ac{MMPF} (Eq.~\ref{eq:mmpf}). For S1,
we follow the approach used in~\citep{2010PhRvD..82j2002N} and select
host galaxies for simulated \ac{BNS} coalesence with each galaxy
weighted by its B-band luminosity based on Eq ~\ref{eq:mmpf}.  We note
that the
\ac{GWGC} is ${\sim}100\%$ and ${\sim}60\%$ complete out to nearly $40$
and $100$Mpc respectively and is estimated using the blue band
luminosity function ignoring the ``Zone of Avoidance",~\citep[see
discussion in ][]{White2011CQGra..28h5016W}.  The effects and an
appropriate consideration of this type of selection bias will be
addressed in future studies. 

We also investigate the effects of using a \ac{MMPF} based on
different galaxy properties. It is not clear which choices of
astrophysical prior have more effect on \ac{GW}-galaxy host studies.
To that end, the B-band luminosity has been often used as a proxy for
the star formation rate of a galaxy. Whether or not the star formation
rate is a good tracer of \ac{CBC} events is much discussed in the
literature~\cite[e.g.][]{2010ApJ...725.1202L,O'Shaughnessy2010ApJ...716..615O,2013ApJ...769...56F,2013ApJ...779...72D}.
Among the observable/inferrable galaxy properties, the stellar mass of
galaxies should be another important (possibly dominating) property to
be used. This is because 1) the \ac{GW} event rate in a galaxy is
expected to be proportional to the total number of stars in that
galaxy, and 2) the star formation rate is positive correlated to the
stellar mass of star-forming galaxies~\citep[e.g.][and references
therein]{2007ApJ...670..156D,2008MNRAS.385..147D,2012ApJ...754L..29W,2013MNRAS.434..423K}.
If the time delay between star-forming and \ac{CBC} mergers is long
enough~\citep[e.g.  some scenarios in
][]{O'Shaughnessy2010ApJ...716..615O}, it is reasonable to believe
that the \ac{GW} \ac{CBC}-\ac{SGRB} rate in a galaxy is proportional
to the old stellar mass of that galaxy, which should be traced by
K-band luminosity.

Unfortunately, the \ac{GWGC} only offers B-band luminosity.  For out
test of \ac{MMPF} priors based on different luminosity measures we
performed simulations using the
\ac{UNGC}~\citep[UNGC,][]{Karachentsev2013AJ....145..101K}, which
contains the most exhaustive list of nearby galaxies properties.  This
catalogue only extends to ${\sim}10$ Mpc and therefore to avoid issues
related to the completeness of the catalogue dominating our results,
we use a network consisting of two initial LIGO detectors and Virgo.
In this scenario the \ac{GW} \acp{SNR} are comparable to those
obtained using sources from the \ac{GWGC} together with the advanced
detector network.  To investigate the effects of using a different
astrophysical prior function, we have performed \ac{BNS} coalescence
injections by selecting host galaxies from the \ac{UNGC} based on
their K-band luminosity in simulations S2 and S3.  We have again made
the assumption that the catalogue is complete however, for S2 and S3
we have ignored the volume of space beyond the range of the catalogue.
We have therefore assumed an \ac{MMPF} that contains only the first
term in Eq.~\ref{eq:mmpf}. The primary aim of these simulations is to
compare the galaxy ranking effects of using astrophysical prior
functions based on the K-band (S2) and B-band (S3) luminosities.
Hence our \ac{MMPF} choice does not influence our results in this
case.
 
In each injection, we generated a \ac{GW} signal from a \ac{BNS}
coalesences using
\emph{lalapps\_inspinj}\footnote{https://www.lsc-group.phys.uwm.edu/daswg/projects/lalsuite.html}
and injected them into simulated noise from detectors at design
sensitivity. Samples from the posterior distribution on
the \ac{GW} parameters are obtained using
lalinference~\cite{2013PhRvD..88f2001A,Veitch2010PhRvD..81f2003V} with
priors on the signal parameters $\GWnon$ being the same as those used
to simulate the signals.  In this case the prior distributions on
$\cos\iota,\psi,\phi_{\text{c}}$ were uniform across their respective
ranges, the priors on $m_{1},m_{2}$ were assumed uniform on the range
$1.3$--$1.5M_{\odot}$ and the time of coalescence was assumed uniform
on the time window $\pm 40$msec around the true injected value.  The
common parameter priors were dealt with according to
Appendix~\ref{sec:GWsamples} with an assumption that the galaxy
catalogue data implicitly incorporates a sky position and distance
prior consistent with a uniform distribution in volume.

The final joint posterior distribution is obtained by computing the
product of the estimated \ac{GW} posterior on the common parameters
and the galaxy catalogue based \ac{EM} prior according to
Eq.~\ref{eq:joint_pos4}. To estimate the \ac{GW} posterior
distribution at the location of a given galaxy using the \ac{GW}
posterior samples, we use a box average method. Namely, we approximate
the probability density at a galaxy location as the ratio of the
number of samples within a 3D box centred at that location to the
total number of samples.  We set the sky location size of each box to
be $\sim 10\%$ of the range covered by all posterior samples (in both
right ascension and declination) and for the box size in distance we
choose $10\%$ of the distance of the galaxy in question. This choice
of density estimation and the choice of box size does affect the final
posterior probabilities at each \ac{GW} host galaxy candidate but does
not strongly affect the relative ranking of density values between
galaxies. We plan to address the method of density estimation from
posterior \ac{GW} samples in future work but note that our approach
here is adequate in terms of this proof-of-principle analysis. 

In general there will be a finite joint posterior contribution from
the region of overlap of the \ac{GW} likelihood and the continuous
portion of the \ac{MMPF} beyond $100$Mpc (for S1).  The integrated
volume of this region represents the joint posterior probability that
the source was associated with an unknown galaxy not included in the
catalogue.  Since such a region is not associated with a specific
galaxy we do not include this region in the ranking (i.e. only known
galaxies are ranked).  However, this probability of the host being
outside the catalogue is included when computing the posterior
probability of each known galaxy (i.e. if there is a $90\%$
probability of the host being outside the catalogue then the highest
possible probability assigned to any known galaxy would be $10\%$).

%
  \begin{deluxetable}{ccccccc}
    \tablecolumns{7} \tablewidth{0pc} \tablecaption{Simulation and
      search parameters\label{simulation_p}} 
    \tablehead{ 
      \colhead{} &
      \colhead{} & 
      \multicolumn{3}{c}{Injection} &
      \colhead{} &
      \colhead{GW search parameters}  \\
      \cline{3-5} \cline{7-7}\\
      \colhead{Simulation} & \colhead{} & \colhead{Galaxy catalogue} & \colhead{Weighting factor} & \colhead{Network} & \colhead{} &
      \colhead{Model}} \startdata
    S1 & & GWGC & B & Adv. L-H-V   & & B \\
    S2 & & UNGC & K & initial L-H-V & & K \\
    S3 & & UNGC & K & initial L-H-V & & B \\
    \enddata
    \tablecomments{For each simulation we indicate the galaxy
      catalogue (either \ac{GWGC} or \ac{UNGC}) and the luminosity-band 
      based weighting factor (B or K) used for simulating sources.
      The network indicates whether the advanced or
      initial \ac{GW} detectors were used and the final column shows
      the model, based on galaxy luminosity, used by our analysis for
      each simulation.  We note that data generated for the S2 and
      S3 simulations were identical with only the analysis differing
      between them.}
  \end{deluxetable}

\section{Results}\label{sec:results}
%
%
\begin{deluxetable}{cc}
  \tablecolumns{1}
  \tablewidth{0pc}
\tablecaption{The S1 case study signal injection parameters\label{tab:case_study_params}}
\tablehead{
    \colhead{Parameter}   & \colhead{Injection value}}
  \startdata
$\alpha$  & 334.7207 deg \\
$\beta$  & -1.0587 deg \\
$d$ &  67 Mpc   \\   
$m_{1}$  & $1.40\,M_{\odot}$ \\
$m_{2}$  & $1.36\,M_{\odot}$ \\
$\mathcal{M}_{\text{c}}$  & $1.20\,M_{\odot}$ \\
$\eta$  & 0.25 \\
$\cos\iota$  & 0.41 \\
$\psi$  & $4.10$ rads \\
\cline{1-2}
Network SNR  & 46.02
\enddata
\end{deluxetable}
%
\begin{deluxetable}{ccccc}
  \tablecolumns{5}
  \tablewidth{0pc}
\tablecaption{Galaxy rankings for the simulated signal plotted in Fig.~\ref{case_study_contours} for different ranking methods\label{case_study_rankings_table}}
\tablehead{
    \colhead{}   & \colhead{} &  \multicolumn{3}{c}{Ranking Method} \\
    \cline{3-5}\\
    \colhead{Ranking} & \colhead{} & \colhead{Galaxy $L_{\text{B}}$} & \colhead{Galaxy $L_{\text{B}}$ and $d$}   & \colhead{GW+MMPF (probability)}}
  \startdata
1 &  &C  &C &  A  (99.36\%)   \\   
2  & &A  &A &  B   (0.37\%) \\   
3 &  &D  &B   &  C  (0.27\%) \\   
4  & &B  &E   & \nodata   \\   
 5 & &E  & \nodata  & \nodata   \\  
\enddata
\tablecomments{Blank entries correspond to galaxies with too low probability to rank under each scheme.}
\end{deluxetable}

\subsection{Case study examples for S1}

\subsubsection{Identifying a \ac{GW} host
  galaxy}\label{sec:case_s1_host}

\begin{figure*}
  \begin{center}
    \includegraphics[width=\textwidth]{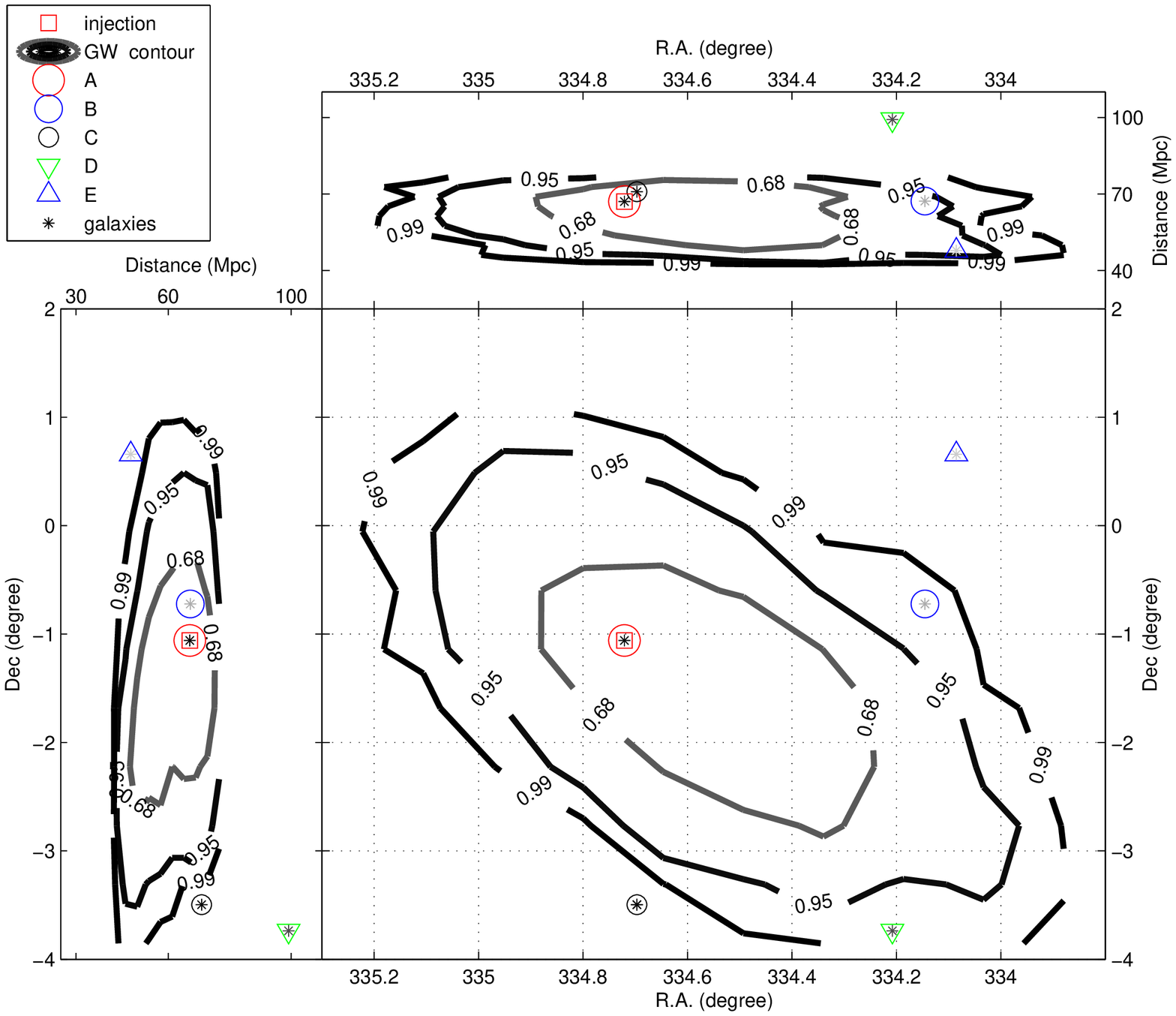}
    \caption{Sky localisation for a single \ac{BNS} coalescence signal
      with injection parameters shown in
      Tab.~\ref{tab:case_study_params} for simulation S1. The contours
      map out the 68\%, 95\% and 99\% confidence regions on the
      estimate of the sky location and distance of the signal
      progenitor obtained using only \ac{GW} observations. Also
      plotted are circle markers corresponding to the first (red
      circle), second (blue circle) and third (black circle) ranked
      host galaxy candidates, labelled A, B and C respectively, as
      determined using Eq.~\ref{eq:joint_pos4}. Galaxy D (green
      downward-pointing triangle) has a lower probability because
      of its distance from the \ac{GW} sky location estimate. Galaxy E
      (upward-pointing triangle) is excluded by Eq.~\ref{eq:joint_pos4} .
      Additionally, grayscale asterisk markers are for all galaxies in
      this sky region, with the shade of the markers corresponding to
      each galaxies B-band luminosity. The darkest markers are the
      most luminous in the B-band. The true sky location and distance
      of the injected \ac{BNS} coalescence signal (red square) in this
      case corresponds to the top ranking galaxy candidate.  The
      simulated signal had an optimal \ac{SNR} of H1: 28.02, L1:
      22.67, V1: 28.60, Network:
      46.02. \label{case_study_contours}}
  \end{center}
\end{figure*}

To illustrate our method, we plot a 12 square degree region around the
injected sky location of a simulated \ac{GW} signal in
Fig.~\ref{case_study_contours}.  Table~\ref{tab:case_study_params}
summarizes the injection parameters for this particular signal. The
galaxies in this region of the sky are plotted with grayscale asterisk
markers with bolder markers representing galaxies with stronger B-band
luminosities. Superimposed on the galaxies are contours bounding the
68\%,95\%, and 99\% confidence regions on the sky location estimate
using only the \ac{GW} data. We note that, by taking the product of
the B-band based \ac{MMPF} (Eq.\ref{eq:mmpf}) with the \ac{GW} sky
location posterior following Eq.~\ref{eq:joint_pos4}, the top ranked
galaxy (A) in this example corresponds to the host galaxy from which
the simulated \ac{GW} signal originates i.e. \emph{the injection host
  galaxy}.  A merit of our approach is that the posterior
probabilities assigned to host galaxy candidates can be used to better
direct further analyses such as providing a refined sky location for
telescopes to observe an \ac{EM} counterpart signal.  In this example,
there are three candidates with non-zero posterior probability with
the first ranked galaxy having a probability of $0.99$ that it is the
\ac{GW} host galaxy.
   
It is interesting to note that galaxy A in
Fig.~\ref{case_study_contours} is only the second most B-band luminous
galaxy in this sky region. In fact, the brightest B-band galaxy has
been ranked third by our analysis. For a detailed comparison,
Table~\ref{case_study_rankings_table} breaks down the rankings of the
five galaxies in this sky region based on three potential ranking
criteria. If we choose to rank galaxies based on B-band luminosity
alone, we obtain rankings shown in the second column.  Similarly, if
we use a statistic that ranks candidate galaxies based on their B-band
luminosity combined with the inverse of galaxy distance, a slightly
different ranking, shown in column 3, is obtained. In this latter case
galaxy D is severely down-weighted due to its distance from the
\ac{GW} sky location estimate. In both of these potential ranking
methods the galaxy from which the \ac{GW} signal originates is not
identified as the top ranked galaxy. On the other hand, by combining
the GW sky location posterior with a B-band based MMPF (column 4), we
identify the host injection galaxy as the top ranked
galaxy. Furthermore, by combining the \ac{GW} sky location posteriors
and the B-band based \ac{MMPF}, two of the five galaxies are not
considered likely host galaxies at all. As before, galaxy D is deemed
too far away from the bulk of the \ac{GW} distance posterior and,
additionally, galaxy E is too remote from the sky location posterior
to be considered significant candidates.

\subsubsection{Enhanced inclination angle
  inference}\label{result_case_incl}

\begin{figure*}
\begin{center}
\includegraphics[width=\textwidth]{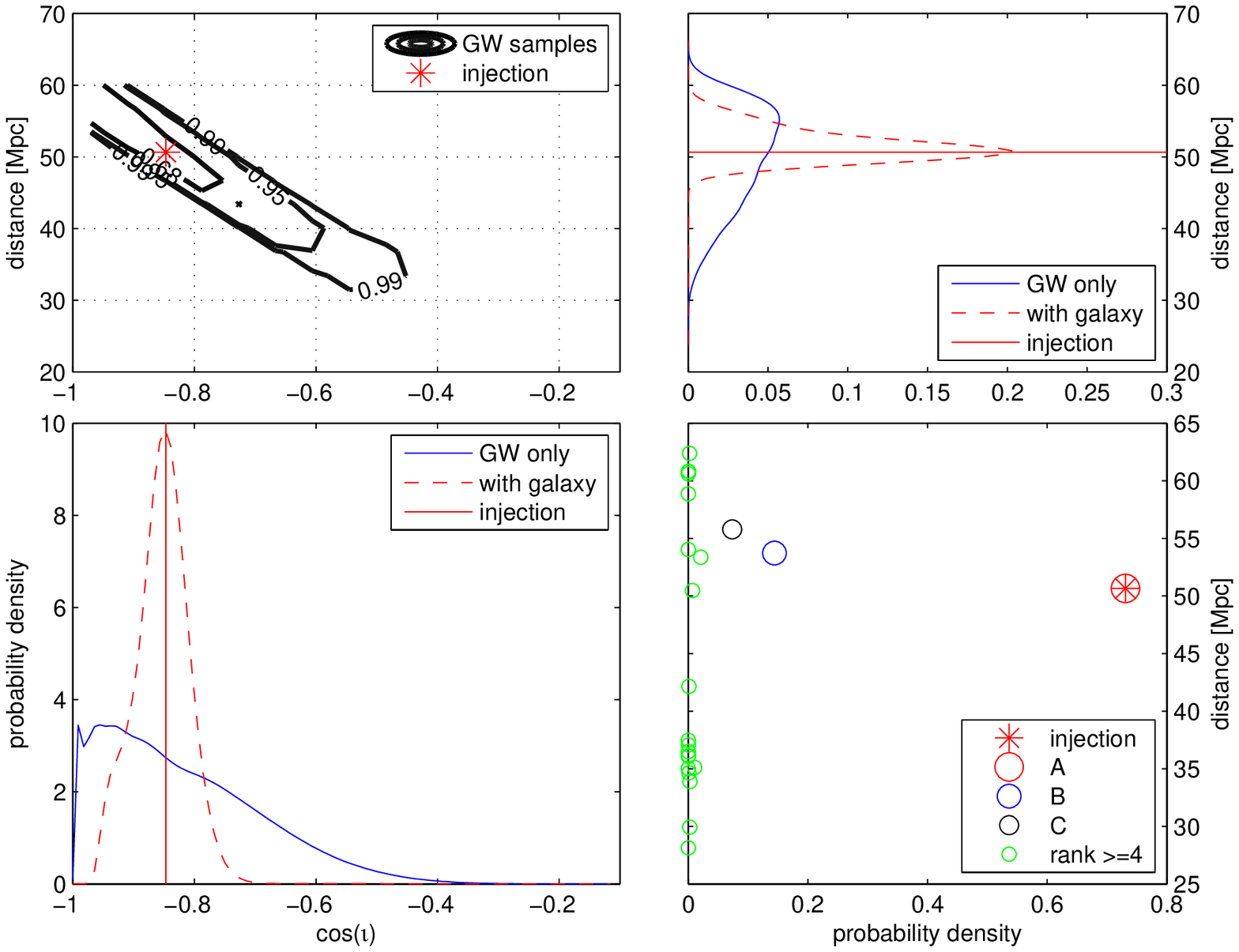}
\caption{An example showing the reduction in the degeneracy between
  distance and inclination angle $\iota$ .  The top-left panel shows
  \ac{GW} posterior samples. The lower-right panel corresponds to the
  marginalised posterior probability of each host galaxy as a function
  of it distance.  The circle markers correspond to the first (red),
  second (blue) and third (black) ranked host galaxy candidates,
  labelled A, B and C respectively, as determined using
  Eq.~\ref{eq:joint_pos4}.  The top-right and lower-left panels
  correspond to the marginlised probability density functions on
  distance and $\cos\iota$, receptively.  The small-sharp features
  seen in the solid blue \ac{GW}-only curve in the lower left-panel
  are artifacts of the smoothing function used to convert samples to
  densities.  Solid blue and red dashed curves correspond to using
  only \ac{GW} posterior samples and \ac{GW}-galaxy catalogue
  information, respectively.  The simulation injection values of
  distance and $\cos\iota$ are shown as solid red lines. The simulated
  signal had an optimal \ac{SNR} of H1: 44.70, L1: 39.32, V1: 27.26,
  Network: 65.48.  \label{case_study_iota1}}
  \end{center}
   \end{figure*}

Once a host galaxy is identified, the distance of the host galaxy can
be used to provide an improved estimate on the inclination angle of
the \ac{CBC} progenitor (see Sec.~\ref{sec:inclination}). With each
galaxy in the \ac{GW} signal error region that is assigned a non-zero
probability, a combined posterior distribution on the inclination
angle can be obtained by multiplying the probably assigned to each
galaxy with the posterior distribution on the inclination angle using
the \ac{GW} signal only (see Eq.~\ref{eq:joint_pos5}). We plot an
example where the inclination angle for the \ac{CBC} system is
improved via our method in Fig.~\ref{case_study_iota1}. We see that
the posterior probability peaks strongly around the correct value when
the injection host galaxy is identified as the top ranking
candidate. Furthermore, the inclination angle posterior distribution
is now much narrower given the additional distance information
provided by the galaxy catalogue.  In this example, the standard
deviation of the $\cos\iota$ posterior is $\pm 0.036$ after combining
the \ac{GW} posterior with the \ac{EM} data.  Using the \ac{GW}
posterior alone gives a much larger value of $\pm 0.12$.

\subsection{Ensemble statistics for S1}

To statistically characterise how well our method, together with the
current galaxy catalog adopted in GW astronomy (the \ac{GWGC}),
correctly identifies the host galaxy, we study the injection host
galaxy rankings and probabilities for the 8000 simulated \ac{GW}
signals.  The 8000 simulations, of which 1000 had hosts selected from
the \ac{GWGC}, produced signals of varying \ac{SNR} since the source
distances ranged up to $200$Mpc, the sensitive range of the advanced
detector network and source orientation parameters were drawn
randomly. Table~\ref{snr_bin} shows the corresponding \ac{SNR}
distribution for simulation S1.

\begin{figure*}
  \begin{center}
    \includegraphics[angle=0,width=3in]{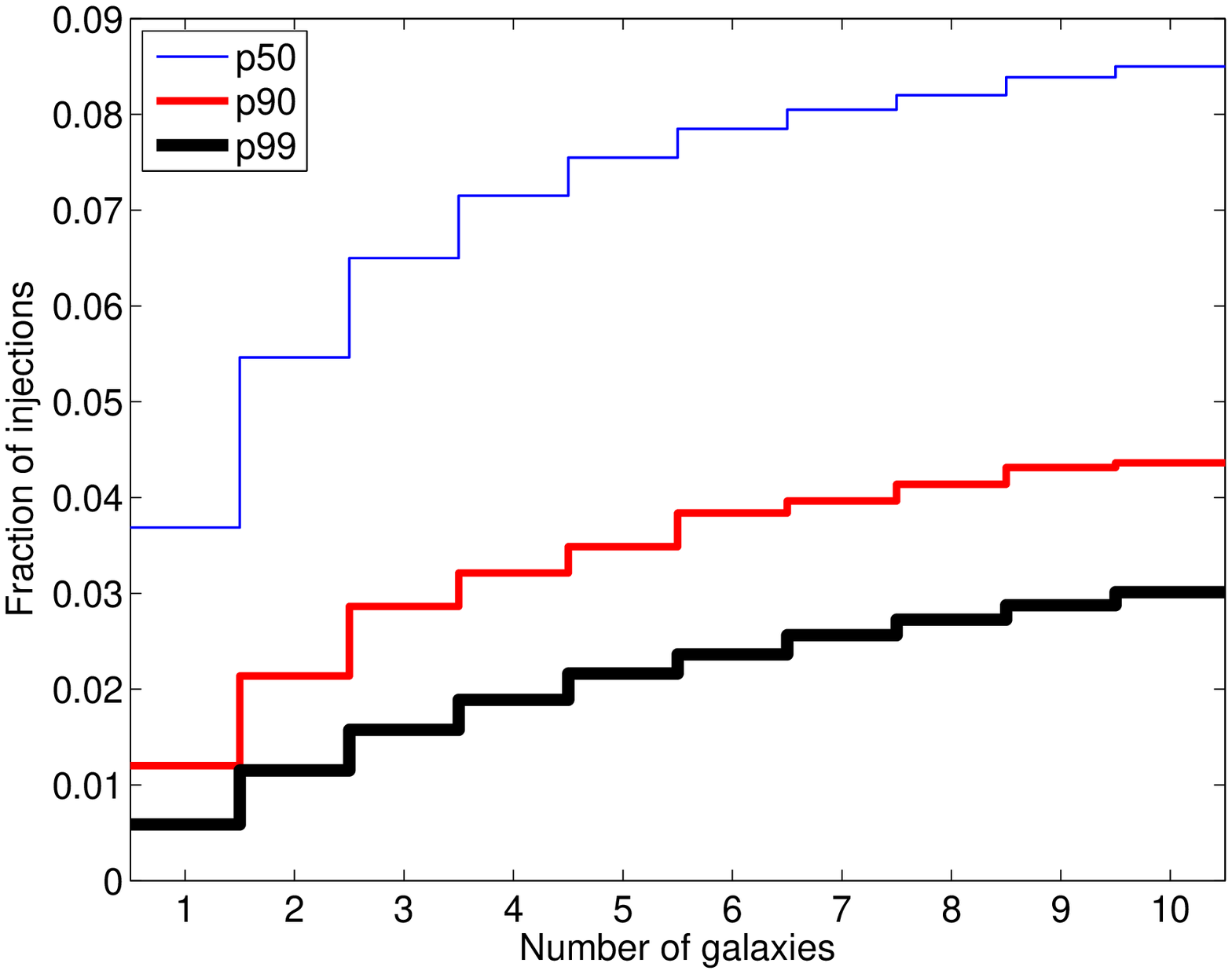}
%
%
    \includegraphics[angle=0,width=3in]{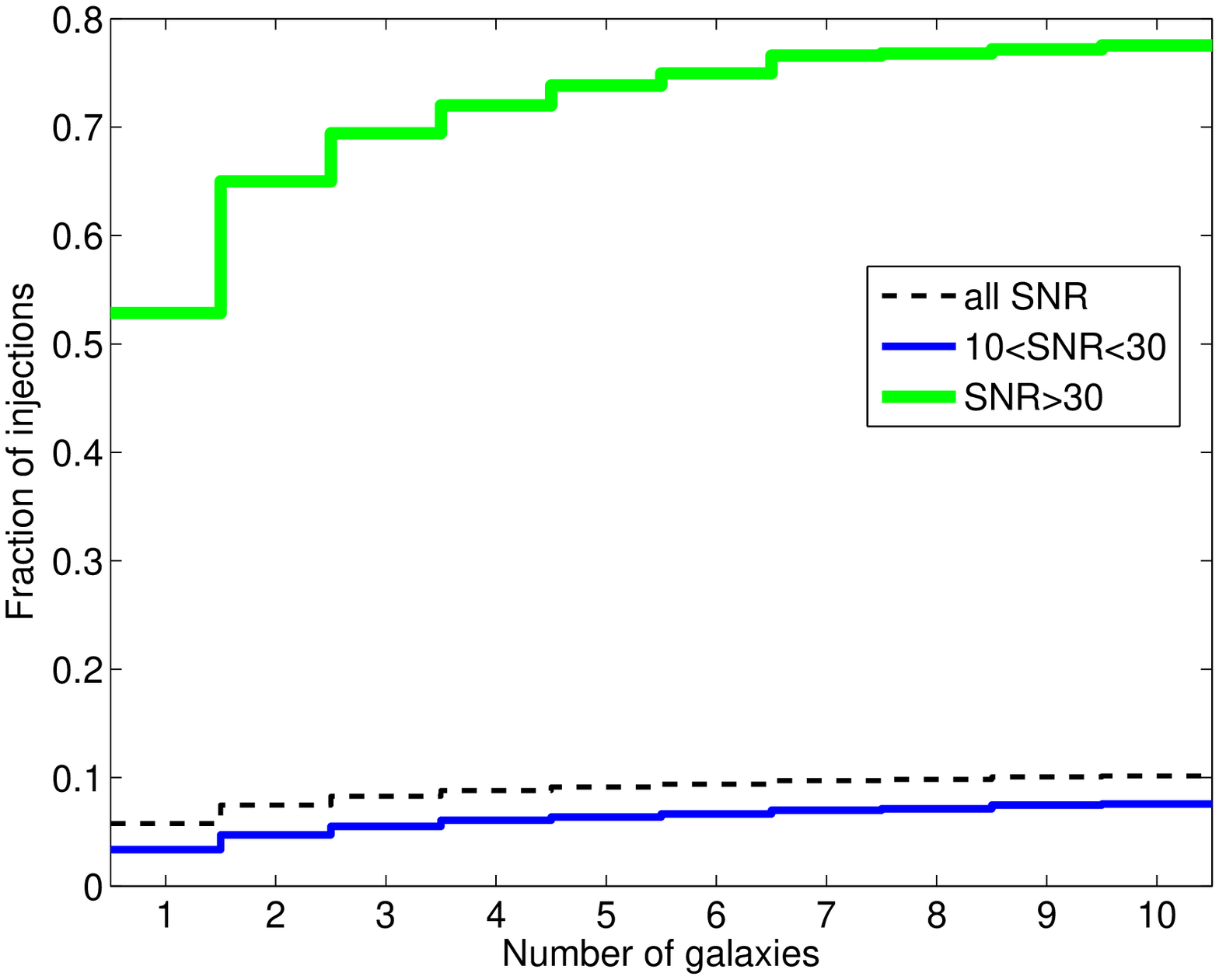}
    \end{center}
    \caption{Left panel: the fraction of analysed signal injections to
      have $50\%$, $90\%$ and $99\%$ probability versus the minimum
      number of galaxies within the \ac{GWGC} required to achieve that
      probability for the S1 simulation.  For example, ${\sim} 4\%$ of
      the time, the top 7 ranked galaxies would contain $90\%$
      of the posterior probability. \label{cp_gwgc_allfree} Right
      panel: the fraction of analysed signal injections to contain the
      true host galaxy at or above a particular ranking as a function
      of that ranking for the S1 simulation.  Different curves are
      shown for various \ac{SNR} ranges.  For example, $10\%$ of the
      time, the true host galaxy is ranked $10$'th or higher when
      considering signals of all
      \acp{SNR}. \label{ranking_inj_snr_s1}}
\end{figure*}

We choose to focus on the following statistical measures to quantify
the effectiveness of our approach.  First we define the minimum number
of galaxies, $N_{Z}$, within the \ac{GWGC} required to cover a
fraction $Z$ of the posterior probability such that
$\sum_{j=1}^{N_{Z}} p_j \geq Z$, where $p_j$ is joint
\ac{GW}-catalogue posterior probability of the $j$'th ranked galaxy
candidate.  In the left panel of Fig.~\ref{cp_gwgc_allfree} we plot
the cumulative fraction of all 8000 S1 simulations required to obtain
a desired probability of $0.5$, $0.9$ and $0.99$ summed over galaxies
against the number of galaxies in the sum. We find that in more than
${\sim}8.5\%$ of all simulations, the top 10 galaxies account for a
total probability of $0.5$. In fact, the single top ranking galaxy
itself has a probability of $0.5$ in ${\sim}4\%$ of all
simulations. Similarly, we see that in ${\sim}4.5\%$ and
${\sim}3\%$ of simulations, the top 10 galaxies sum up to
probabilities of $0.9$ and $0.99$ respectively.

 %
\begin{deluxetable}{cccc}
  \tablewidth{0pt} \tablecolumns{4} \tablecaption{fraction in optimal network
    SNR bins for each simulation\label{snr_bin} }
      \tablehead{     &  &  fraction     \\
SNR bins    &0-10&10-30&$>$30} 
  \startdata
  S1  &0.295 &0.637 &  0.068       \\
  S2 (S3) &       0.149    &   0.586   & 0.265          \\
\enddata
\end{deluxetable}
%
%
%
As an alternative figure of merit, we also consider the number of
galaxies within the \ac{GWGC}, ordered by their ranking, required
before the true \ac{GW} injection host galaxy is included.  As is
shown in the right panel of Fig.~\ref{ranking_inj_snr_s1},
${\sim}5\%$ of all injected signals are correctly identified as the
highest ranking.  We must stress that only $12.5\%$ of injections in
total had a host galaxy from the \ac{GWGC} catalogue with the
remainder being drawn from a continuous distribution, uniform in
volume, beyond $100$Mpc. Hence the majority of all injections for
which no host galaxy was identified were due to them not being drawn
from the \ac{GWGC}.  If we focus on ``loud'' events, such as events
with \ac{SNR} $>30$, we find that there is a ${\sim}70\%$
probability that we find the true \ac{GW} host galaxy within the top 3
ranked galaxies.  For quieter events, such as those with $10<$ SNR
$<30$, there is ${\sim}8\%$ chance that the top 10 galaxies would
include the \ac{GW} host galaxy.  For very low SNR ($<10$) events,
there is almost no chance ($\sim 0.1\%$), that
the highest ranked galaxy is the true host of the \ac{GW} source.

\begin{figure*}
\begin{center}
  \includegraphics[width=\textwidth]{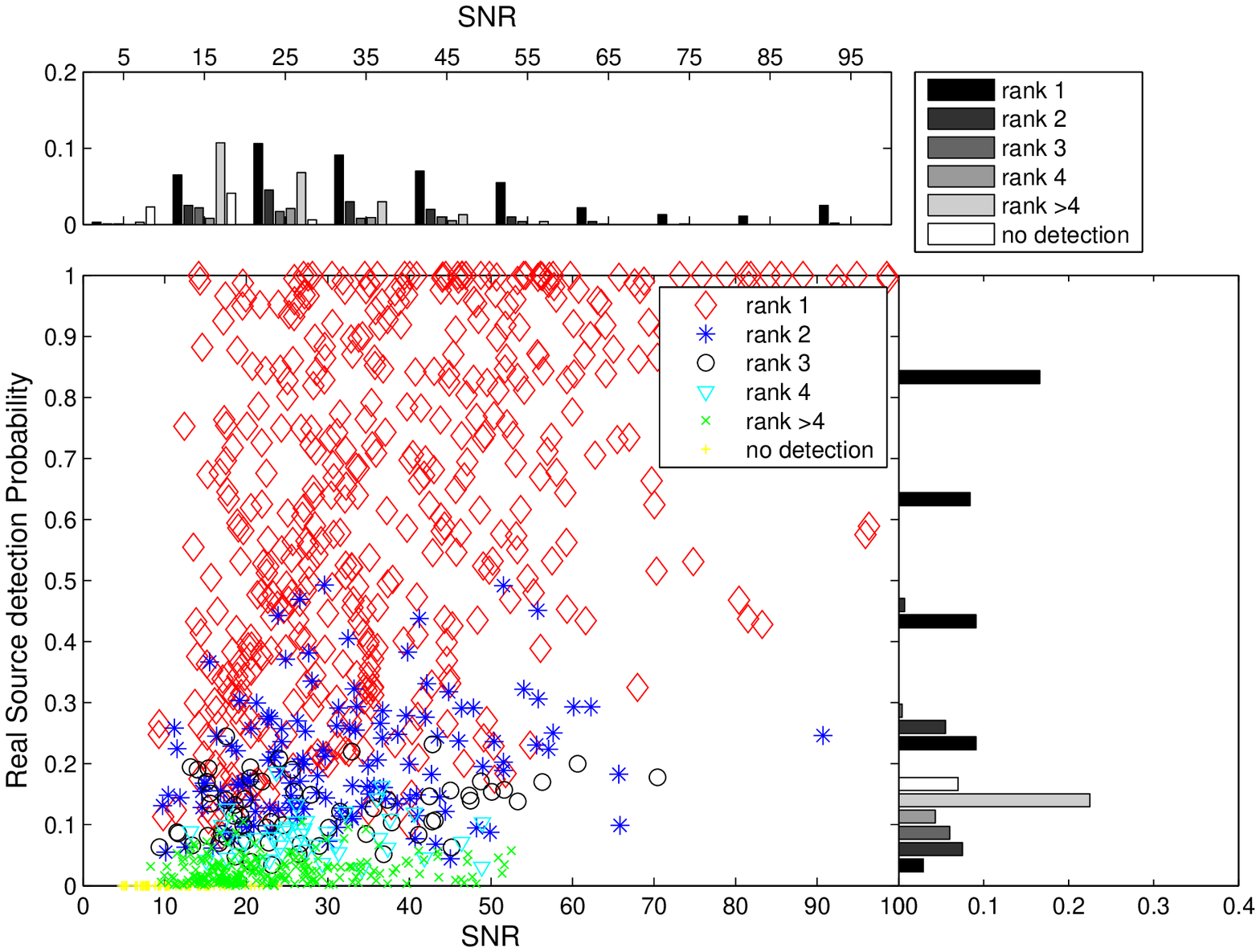}
  \caption{A scatter plot showing the probability of the combined
    \ac{GW}-galaxy catalogue signal posterior versus the injected
    \ac{SNR} for the 1000 simulation in S1 that originated from a
    \ac{GWGC} galaxy.  Plotted above the central scatter plot are the
    fractions of all 1000 injections per rank for various \ac{SNR}
    ranges. A similar histogram is plotted to the right of the central
    plot showing the rank distribution for various ranges in injection
    host galaxy probability.  We note that most simulations originating from
    galaxies in the \ac{GWGC} catalogue result in signals with an optimal network \ac{SNR}
    greater than 10. For these simulations, all galaxies with a
    probability of 0.2 or more are ranked top 3 or better.  Some
    injection host galaxies for high \ac{SNR} signals are not
    identified as the top ranking galaxy due to a combination of
    nearby galaxies with greater B-band luminosities and \ac{GW} sky
    location posteriors being peaked away from the injection host
    galaxy. About $10\%$ of all simulations are classified {\it no
      detection} because a posterior probability was not assigned to
    the injection host galaxy since no \ac{GW} posterior samples were
    found within our galaxy-centred sample boxes.  Almost all such
    simulations had signals with \ac{SNR} less than 20. \label{prob_snr_scatter_adv_lb}}
\end{center}
\end{figure*}

To investigate the efficiency of our method at identifying host galaxies, 
we examine the behavior of simulations that originate from a galaxy
in \ac{GWGC}. In Fig.~\ref{prob_snr_scatter_adv_lb}, the central
scatter plot shows the \ac{SNR} versus the posterior
probability obtained for the true host galaxy of the injected
signal in . We stress that each point on the plot corresponds to these values for each of
the 1000 \ac{GWGC} based S1 simulations and we do not include events
injected from beyond the \ac{GWGC}. To the top and left of the central
plot are multiple histograms showing the distribution of the \acp{SNR}
and injection host galaxy probability for various injection host
galaxy ranks.

For optimal network \ac{SNR} $>20$ injections, the top ranking of the
injection host galaxy has only a weak dependence on the \ac{SNR} (see
the ranking depended \ac{SNR} histogram in
Fig.~\ref{prob_snr_scatter_adv_lb}).  However, some injection host
galaxies for high \ac{SNR} events are not ranked first due to the
presence of nearby galaxies with greater B-band luminosities or the
\ac{GW} sky location posterior being peaked away from (but still
consistent with) the injection host galaxy. Furthermore, we note that
approximately $10\%$ of the injections were not assigned a posterior
probability since these sources yielded no \ac{GW} posterior samples
within our galaxy-centred sample boxes used for density estimation.
We classify these injections as being particularly poorly localised
via the \ac{GW} observation and have been identified as due primarily
to low \ac{SNR} injections as indicated by the yellow (cross) markers in the
scatter plot in Fig.~\ref{prob_snr_scatter_adv_lb}.

\subsection{Using a different MMPF}

In the simulation described above the sky location of the injected
\ac{CBC} signals were injected \emph{and} recovered using our approach
with a B-band luminosity based \ac{MMPF}.  In simulations S2 and S3
injections were performed using K-band luminosities to determine the
relative galaxy \ac{MMPF}.  In simulation S2 the same K-band based
\ac{MMPF} was used for recovery and in simulation S3 the effects of
using the B-band for recovery have been studied.  In both simulations
the actual injection host galaxies were identical, i.e. the same
realisation.  These 2 simulations have been performed using the
initial LIGO-Virgo detector network at design sensitivity since, of
our chosen catalogues, the \ac{UNGC} contains both B and K-band
information but has a range of only $\sim10$ Mpc.

The colour index, defined by $<\text{B}-\text{K}>$, depends on galaxy
morphology~\citep[e.g.][]{Jarrett2003AJ....125..525J}.  Therefore, for
K-band luminosity based injections, B-band luminosity together with
galaxy morphology could be used as a proxy for the K-band to recover
the injection within our \ac{MMPF}.  However, the details of such an
analysis are the subject of ongoing study within the multi-messenger
astronomy community. Instead, we use simulation S3 to show the effect
of using B-band luminosity information for signals simulated using
K-band information within our \ac{MMPF}.

\subsubsection{Case study example}\label{result:case:kb}

\begin{figure*}
 \begin{center}
\includegraphics[angle=0,width=3in]{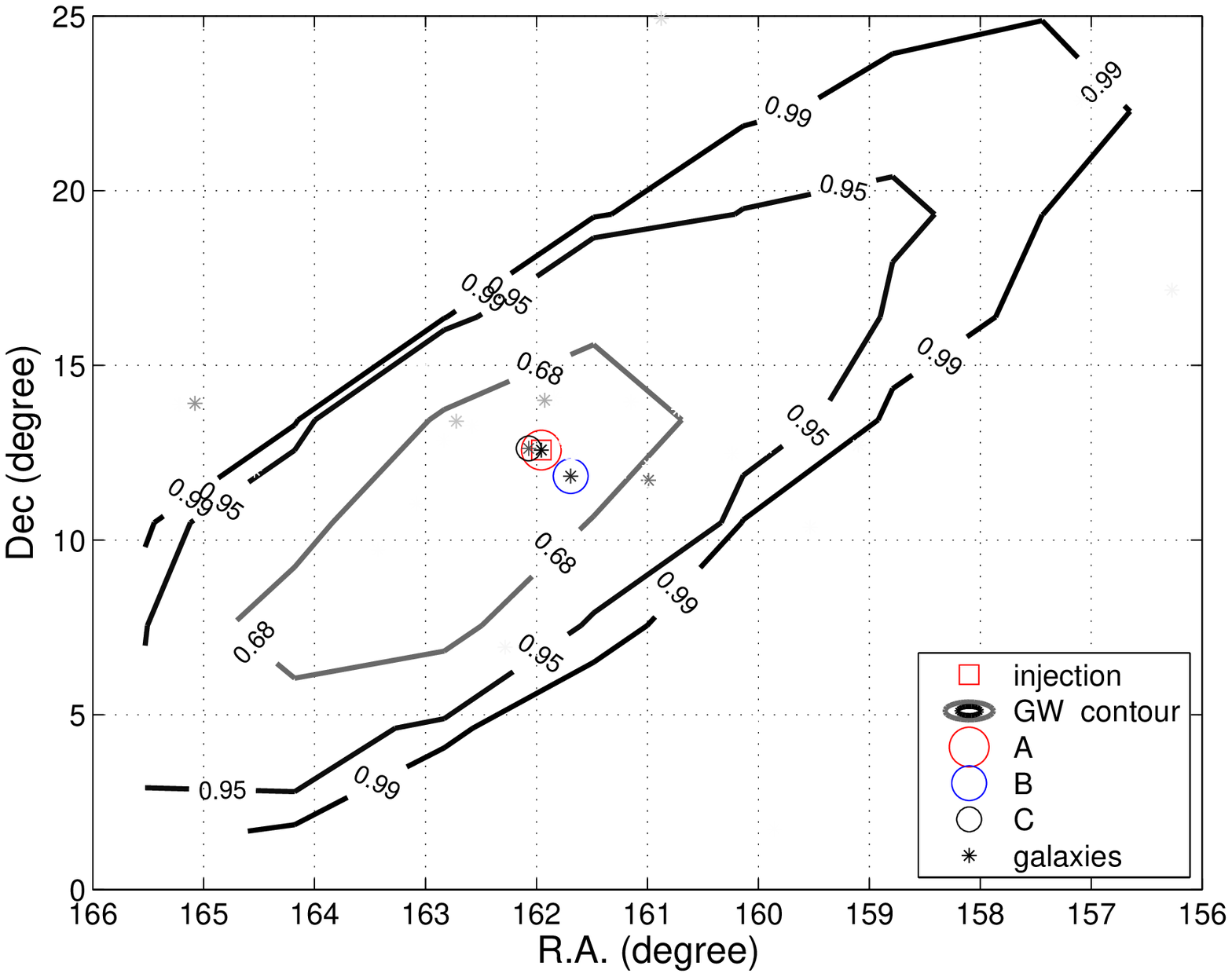}
\includegraphics[angle=0,width=3in]{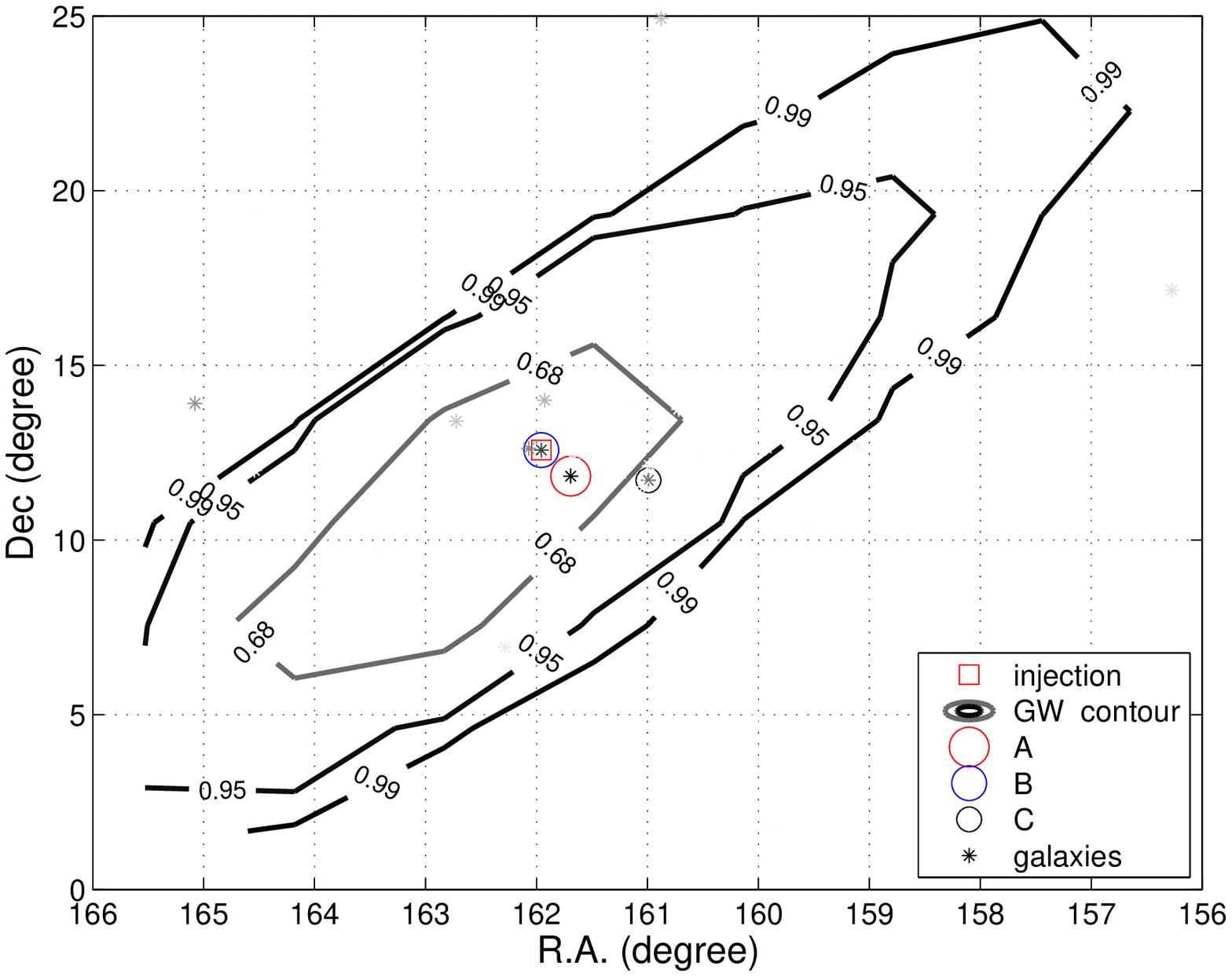}
    \end{center}
\caption{Comparison of the first three \ac{GW} host galaxies in rank
  recovered using in a K-band based \ac{MMPF} in simulation S2 (left panel),
  and in  a B-band based \ac{MMPF} in simulation S3 (right panel).  Symbols have
  the same meaning as defined in Fig.~\ref{case_study_contours}. The
  simulated signal has an \ac{SNR} of H1: 8.19, L1: 9.92, V1: 6.91 ,
  Network: 14.60.  \label{one_injection_bk_fig}}
  \end{figure*}

Using the ``correct" \ac{MMPF} should lead to a better rank for the
injection host galaxy when considering a population of sources.  In
this section we give an example where using the correct \ac{MMPF}, by
which we mean the same \ac{MMPF} as that used for simulating the
\ac{GW} signal, leads to a better ranking for the injection host
galaxy (see Fig.~\ref{one_injection_bk_fig}). In particular, the
example K-band injection we have chosen ranks as the first in S2, our
 K-band based recovery \ac{MMPF} (shown in left panel of 
Fig.~\ref{one_injection_bk_fig}) and ranks second when using the
B-band based recovery \ac{MMPF} (shown in right panel
Fig.~\ref{one_injection_bk_fig}).
We also note that the 3 highest ranking galaxies in both
simulations only share 2 common galaxies.
The third rank galaxy
estimated using the K-band based recovery \ac{MMPF} lies within the
distance marginalised \ac{GW} posterior 1-$\sigma$ sky map contour,
while the third ranking galaxy using the B-band based recovery
\ac{MMPF} lies outside the \ac{GW} posterior 1-$\sigma$ sky map
contour. The sky separation of the first and the third candidates
returned by the B-band based recovery \ac{MMPF} (the ``non-correct"
one) may challenge an \ac{EM} follow-up team in terms of observing the
top 3 candidates given the limited field of view of \ac{EM}
telescopes.
 
\subsubsection{Ensemble statistics}

\begin{figure*}
\begin{center}
\includegraphics[angle=0,width=3in]{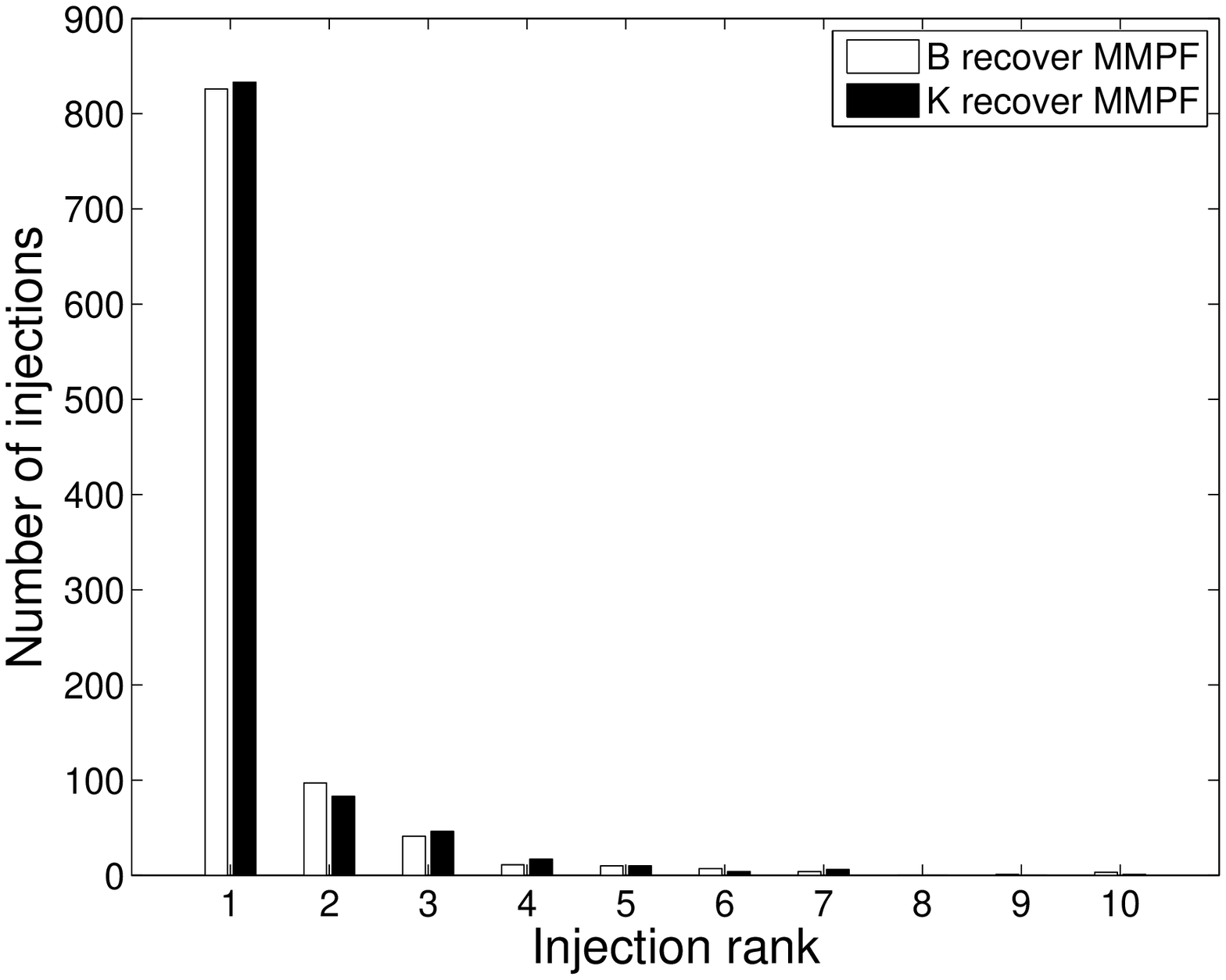}
\end{center}
 \caption{Comparison of \ac{GW} host galaxy rank in  S2 and
 S3 simulations.  The  K-band  based recover MMPF adopted  in S2 and B-band  based recover MMPF adopted  in S3 return similar statistical results on injection rank. \label{kvb_hist_rank}}
\end{figure*}

Distributions of the ranking of the injection host galaxy are shown in
Fig.~\ref{kvb_hist_rank}. There are 22 of 1000 injections that have a
different rank depending on whether the recovery \ac{MMPF} is a
function of the K or B-band luminosity.  The K-band based recovery
\ac{MMPF} adopted in S2 and B-band based recover \ac{MMPF} adopted in
S3 return similar statistical results on injection rank (
Fig.~\ref{kvb_hist_rank}). 
Therefore, for this particular galaxy catalog, one can
safely use B-band to construct the \ac{MMPF} when K-band data is not
available, although, as one would expect, for some particular cases
the K and B-band based \ac{MMPF} show different results (e.g. see
Fig.~\ref{one_injection_bk_fig}).

\section{Conclusions and discussion}\label{sec:discussion}
We have constructed a Bayesian approach to multi-messenger astronomy
and described a proof-of-principle analysis for this approach.  The
analysis chosen in this case was designed for joint \ac{EM} and
\ac{GW} observations, in particular, galaxy catalogues and \ac{GW}
events from \acp{CBC}.  The aim of this research is to improve the sky
localisation of \ac{GW} events by identifying the \ac{GW} source host
galaxy.  Identifying the \ac{GW} host galaxy is, for example, a vital
component in using \ac{GW} signals as cosmological standard-sirens
\citep[e.g.][]{Schutz1986Natur.323..310S,2012PhRvD..86d3011D} as well
as a key ingredient for follow-up observations for \ac{EM}
counterparts to \ac{GW} signals
\citep[e.g.][]{2008CQGra..25r4034K,Veitch2010PhRvD..81f2003V,2013ApJ...767..124N}.
The proof-of-principle analysis presented here has been demonstrated
using simulated \ac{BNS} events, with their sky locations and
distances randomly selected from the \ac{GWGC} and uniform distribution for number density  (in S1) and \ac{UNGC} (in S2 and  S3). 
Additional prior galaxy weighting was based on individual galaxy
measured/inferred parameters and simulated \ac{GW} signal injections
were added to LIGO-Virgo simulated noise. To ensure the \ac{SNR}
distributions are sensible and comparable between choices of
catalogue, we have adopted advanced and initial LIGO-Virgo design
noise performance when using the \ac{GWGC} and \ac{UNGC} related
simulations, respectively.

In addition to the standard galaxy candidate ranking, one merit of our
approach is that the posterior probability of hosting the \ac{GW}
source associated with each galaxy is also produced.  This information
could, for example, be used to better guide \ac{EM} follow-up
observations to focus on the particular galaxies with high posterior
probability (see case study in Sec.~\ref{sec:case_s1_host}) or to
simply limit the galaxies followed-up to those accounting for the bulk
of the total probability (see case study in
Sec.~\ref{result:case:kb}).  Moreover, we have shown that better
constraints on the common parameters of the \ac{EM}--\ac{GW}
observations can enhance the inference on non-common \ac{GW}
parameters.  A case study on the improvement on the inference of the
inclination angle of \ac{CBC} events was present in
Sec.~\ref{result_case_incl}.

Using 8000 S1 simulations, we found that about $8\%$, $4\%$ and $3\%$
of injections have $50\%$, $90\%$ and $99\%$ of the probability
included in the top 10 ranked galaxies in \ac{GWGC}, respectively.
The first ranking galaxy has a $50\%$ probability of being the true
\ac{GW} host galaxy in about $4\%$ of injections.  These result are
dominated by the \ac{GWGC} 100 Mpc distance cut comparing with the
expected reach of Advanced LIGO and Advanced Virgo at design
sensitivity $\sim$ 200 Mpc. A deeper and all-sky galaxy catalog is
necessary to improve the identify of GW host galaxy.

Although for some particular cases the ``correct'' K-band based
recovery \ac{MMPF} shows better results (e.g. see case study in
Sec.~\ref{result:case:kb}), the K and B-band based recovery \ac{MMPF}
return statistically similar results on injection
rank 
for 1000 injections with a K-band based \ac{MMPF} using the
\ac{UNGC}. This may result from the fact that 1) the red-sequence
galaxies represent only $17\%$ of the total number of galaxies in the
local Universe and, 2) it is rare to have a massive red-sequence (no
or little ongoing star-formation) galaxy in close proximity to a
massive blue cloud (star-forming) galaxy.

We note that the proposed method for incorporating astrophysical
information is very flexible. It is straightforward to incorporate a different 
or updated galaxy catalogue into the \ac{MMPF} by updating the
sky location and distance parameters in Eq.~\ref{eq:mmpf1}. A new, more
sophisticated \ac{MMPF} can also taken into account by 
updating Eq.~\ref{eq:mmpf2}. In both cases, there is no need
to perform the potentially time consuming step of re-analysing 
the \ac{GW} data since the \ac{MMPF} is constructed
to be independent of \ac{GW} observations. It is just a matter of 
constructed the \ac{MMPF} using Eq.~\ref{eq:mmpf} and multiplying
the new \ac{MMPF} with the \ac{GW} posteriors.

Faint galaxies are a challenge for any astronomical survey and
therefore also an obstacle in identifying a \ac{GW} host galaxy using
any available galaxy catalogue.  The completeness of a catalog is a
complex concept and not particularly well defined.  Analysis of
luminosity functions can give an indication of the level of
incompleteness in a catalogue.  In~\cite{White2011CQGra..28h5016W}
they estimate the completeness of the \ac{GWGC} as a function of
distance by comparing blue band data with a analytical Schechter
galaxy luminosity function within 100 Mpc. 
In this paper,  we only take into account  the effect of  galaxy catalog distance cut  on  identifying  GW host galaxy in Eq.~\ref{eq:mmpf1} , which is the  major  task to identify GW host galaxy by \ac{GWGC} in advanced detectors era.    Further effects by the  \ac{EM} data, such as the  completeness  of the catalog  within distance cut and  the uncertainty of  \ac{EM} measurements, will be studied in our future work.

One of major components of our approach is the \ac{MMPF}, within which
we distill our astrophysical knowledge related to our underlying
model in Eq.~\ref{eq:mmpf2}, that our \ac{GW} source resides within a galaxy.  In past
searches, the B-band luminosity of galaxies is adopted within the
\ac{GW} community to estimate the \ac{CBC} event rate in galaxies for
proposed galaxy catalog based \ac{EM} follow-up observations
\citep[e.g.][]{2010PhRvD..82j2002N,Evans2012ApJS..203...28E,2014ApJS..211....7A}.
The nature of compact binaries suggests that \acp{GW} from \acp{CBC}
would most likely hosted by the older stellar population in more
massive galaxies. Stellar masses are mainly determined by observed
stellar light through the stellar mass-to-light ratio or fitting the
spectral energy distribution of galaxies
\citep[e.g.][]{2013ApJ...768..178F}, which varies according to a few
parameters~\citep[see a recent review][and references
therein]{CourteauRevModPhys.86.47}.  Therefore, multi-band luminosity
could benefit \ac{GW}-galaxy host research.  Debate pertaining to the
reliability of optical and near-infrared stellar mass estimates is
currently ongoing. However, the fact that the stellar mass-to-light
ratio varies less in near-infrared bands than in blue bands over a
wide range of star-formation
history~\citep[e.g.][]{Bell2001ApJ...550..212B}, and old stellar
populations ($\geq 2$Gyr) are mostly bright in the near-infrared
band~\citep[e.g.][]{Maraston1998MNRAS.300..872M}, suggests that the
near-infrared band (e.g. the K-band) is a better tracer of old
stellar-population mass, and therefore the \ac{CBC} derived \ac{GW}
event rate.  It is believed that the morphology and metallicity of a
galaxy will affect its \ac{CBC} event rate
~\citep[e.g.][]{Belczynski2010ApJ...715L.138B,
  Fryer2012ApJ...749...91F,O'Shaughnessy2010ApJ...716..615O} and its
\ac{LGRB} (possibly associated with \ac{GW} bursts) event rate
\citep[e.g.][]{Fan2010A&A...521A..73F}.  We note that our observations
from using \ac{UNGC} are dominated by the very low galaxy density at
10\,Mpc. Therefore, the selection effects arising from this low
density must be taken into account which, as previously mentioned, is
the scope for future work.  Nonetheless, just as we have started to do
with B and K-band luminosities, it is important for future
multi-messenger analyses to investigate the influence of this
information and/or lack thereof in their simulations.
    
The density of galaxies around the injection host galaxy and the size
of the \ac{GW} posterior on the signal sky location have a significant
impact on the galaxy rankings we have observed. Therefore the galaxy
environment may also play a role on identifying \ac{GW} host galaxies.
Galaxy clusters usually have a massive galaxy surrounding less massive
galaxies.  It will be interesting in the future to test the efficiency
of \ac{GW} host galaxy identification for different \ac{GW} sources
such as \acp{LGRB} associated \ac{GW} burst sources, which are
believed to be preferentially hosted by faint (irregular) galaxies
\citep[e.g.][]{2012grbu.book..269F,Fan2010A&A...521A..73F}.

Beyond our underlying model, that the \ac{GW} source resides within a
galaxy, the potential offset in the observed a \ac{GW} signal from the
centre of its host galaxy (e.g.  by supernovae kicks) has been
suggested by population models~\citep[e.g.][]{1999MNRAS.305..763B}
and observed~\ac{SGRB}
offsets~\citep[e.g.][]{2013ApJ...776...18F}. This effect could be
accounted for using a model that assigns a distribution to common
parameters (sky location and distance) covering the offsets.  The
potential offsets, which are $\sim$ a few to tens of Kpc, should have
minimal effect on the \ac{GW} host galaxy identification even with the
lack of faint galaxies in the catalog.  This is attributable to the
fact that the offset \acp{SGRB} are unlikely to be hosted by the
unobserved faint galaxies which are far from the \ac{SGRB}
hosts~\citep[e.g.][]{2014arXiv1401.7851B}.

Amongst other dependencies, the potential implications of our Bayesian
approach are a function of the common parameters between the two
different observation sets.  There is much discussion regarding the
possible non-\ac{GW} signatures of \ac{BNS} mergers~\citep[e.g.][among
others]{2012ApJ...746...48M,2013PhRvD..88d3010G} and \acp{BNS} are
commonly accepted as the central engine for
\acp{SGRB}~\citep[e.g.][]{1986ApJ...308L..43P,1989Natur.340..126E,2005Natur.437..845F}. The
opportunity to perform multi-messenger astronomy by observing the
\ac{SGRB} counterpart to the \acp{GW} emitted is one of the many
reasons that \ac{CBC} systems are considered an interesting
source. Furthermore, weaker kilonova optical counterparts are also
expected to be emitted by
\acp{CBC}~\citep[e.g.][]{2010MNRAS.406.2650M}.  While \acp{SGRB} are
expected to be highly beamed, the kilonova signal radiates
isotropically.  Direct detection of \acp{GW} in coincidence with their
\ac{SGRB} or kilonova counterparts will provide the strongest evidence
that \ac{SGRB} progenitors are merging \ac{CBC} systems. Besides the
sky location and distance (if available), the common parameters
between \ac{EM} and \ac{GW} signals in these cases could also include
the arrive time and source energy which would then be incorporated
consistently into the \ac{MMPF} design.

Precise sky localisation and the consequent host galaxy identification
is of prime importance to the most promising and well established
ideas in \ac{GW} cosmology.  In~\citep{Schutz1986Natur.323..310S}
and~\citep{2012PhRvD..86d3011D} the idea was proposed and investigated
that correctly combining the potential host galaxy redshifts with the
luminosity distance inferred from \ac{GW} observations would allow
measurement of the Hubble constant using first and second generation
\ac{GW} detectors. Improved host galaxy identification such as the
method we propose would directly impact and reduce the statistical
noise inherent to this cosmological measurement.

As a final remark we consider the imminent \ac{GW} detection era and
the potential 10s--100s of \ac{CBC} signals detectable with the
advanced network of detectors.  In such a scenario our approach could
easily be inverted to ask a different question, what is the true
\ac{MMPF}?  With multiple \ac{GW} detections it would then be possible
to perform model selection on different choices of \ac{MMPF} allowing
the \ac{GW} data to feedback population information to the wider
astrophysical community.

\acknowledgments

We would like to acknowledge valuable input from J.~Kanner, M.~Hendry,
P.~Raffai, and our anonymous referee whose input has greatly improved
the manuscript. The authors also gratefully acknowledge the support of this
research by the Royal Society, the Scottish Funding Council, the
Scottish Universities Physics Alliance and and the Science and
Technology Facilities Council of theUnited Kingdom.  XF acknowledges
financial support from National Natural Science Foundation of China
(grant No.~11303009).  XF is a Newton Fellow supported by the Royal
Society and CM is a Lord Kelvin Adam Smith supported by the University
of Glasgow.

\appendix

\section{Generating \ac{GW} posterior samples}\label{sec:GWsamples}

In this section we describe the practical procedure used to generate
samples from the \ac{GW} posterior $p(\com|\GWdat,M,I)$.  In order to
avoid the step of dividing this distribution by the prior on the
common parameters (see Eq.~\ref{eq:joint_pos4}) we instead effectively
sample from the ratio of posterior and the prior.  Using Bayes theorem
and the assumption that the joint prior on the common and non-common
\ac{GW} parameters is separable such that
$p(\com,\GWnon|I)=p(\com|I)p(\GWnon|I)$ we can express this ratio as
\begin{equation}
  \frac{p(\com,\GWnon|\GWdat,M,I)}{p(\com|I)}=\frac{p(\GWdat|\com,\GWnon,M,I)p(\GWnon|I)}{p(\GWdat|I)}
\end{equation}
This assumption is valid for our \ac{GW}-galaxy catalogue simulations
where the sky position and distance priors are independent of the mass
and orientation parameters of the \ac{CBC}.

Our expression now has explicit priors on $\GWnon$ but in practice, in
order to generate a posterior distribution using existing algorithms,
we are required to specify a prior on all parameters within the
problem, including $\com$ .  In the \ac{GW} analysis case this means
specifying uniform dummy priors on $\com$ such that the function from
which samples are drawn is actually
\begin{equation}
  X(\com,\GWnon)=\frac{1}{\mathcal{V}_{\com}}
  p(\GWdat|\com,\GWnon,M,I)p(\GWnon|I)
\end{equation}
where $\mathcal{V}_{\com}$ is the volume of the common parameter space
and its inverse is the uniform prior on $\com$.  Therefore we specify
non-standard priors for the sky position and distance within our
sampling algorithm since we assume that the \ac{EM} observation
already contains these priors.

Substituting this into Eq.~\ref{eq:joint_pos3}  gives us
\begin{equation}
p(\com,\GWnon|\GWdat,\EMdat,M,I) =\frac{p(\GWdat|I)}{p(\GWdat|\EMdat,M,I)}p(\com|\EMdat,M,I)\mathcal{V}_{\com}X(\com,\GWnon).
\end{equation}
From this point we can proceed as described in
Secs.~\ref{sec:combining} and~\ref{sec:inclination} whereby terms in
the joint \ac{EM}-\ac{GW} posterior are marginalised over a subset or
all of the non-common parameters.  We have therefore made sure that
the correct physical priors on the non-common parameters have been
applied and we have not over-applied the common parameter priors.

\bibliographystyle{apj}
\bibliography{Bibliography}
\end{document}